\documentclass[12pt,preprint]{aastex}
\usepackage{amsmath}
\usepackage{amssymb}

\newcounter{qub}
\newcommand{\qq}{\addtocounter{qub}{1}\arabic{qub}}

\newcommand{\MA}{M31}

\shorttitle{A Search for Planetary Nebulae in Nearby Galaxies With SDSS}
\shortauthors{Kniazev et al.}

\begin{document}

\title{
A Search for Planetary Nebulae With the SDSS: the outer regions of \MA\altaffilmark{0}
}

\author{Alexei Y.\ Kniazev\altaffilmark{1,2,3,4,7},
Eva K.\ Grebel\altaffilmark{4},
Daniel B.\ Zucker\altaffilmark{5,6},
Hans-Walter Rix\altaffilmark{7},
David Mart\'{i}nez-Delgado\altaffilmark{4},
Stephanie A.\ Snedden\altaffilmark{8}
}

\email{akniazev@saao.ac.za}

\altaffiltext{0}{
Based in part on observations collected at the German-Spanish 
Astronomical Center (DSAZ),
Calar Alto, operated by the Max-Planck-Institut f\"{u}r
Astronomie Heidelberg jointly with the Spanish National Commission 
for Astronomy}
\altaffiltext{1}{South African Astronomical Observatory, 
PO Box 9, 7935, South Africa}
\altaffiltext{2}{Southern African Large Telescope Foundation, 
PO Box 9, 7935, South Africa}
\altaffiltext{3}{Sternberg Astronomical Institute,
Lomonosov Moscow State University, Moscow, Russia}
\altaffiltext{4}{Astronomisches Rechen-Institut, Zentrum f\"ur Astronomie
der Universit\"at Heidelberg, M\"onchhofstr.\ 12 -- 14, 
D-69120 Heidelberg, Germany}
\altaffiltext{5}{Department of Physics and Astronomy,
Macquarie University, NSW 2109, Australia}
\altaffiltext{6}{Australian Astronomical Observatory, P.O. Box 296,
Epping, NSW 1710, Australia}
\altaffiltext{7}{Max-Planck-Institut f\"ur Astronomie, K\"onigstuhl 17,
      D-69117 Heidelberg, Germany}
\altaffiltext{8}{Department of Astronomy, New Mexico State University,
Box 30001, Las Cruces, NM 880033, USA}

\begin{abstract}
We have developed a method to identify planetary nebula (PN) candidates in
imaging data of the Sloan Digital Sky Survey (SDSS). This method
exploits the SDSS' five-band sampling of emission lines in PN spectra,
which results in a color signature distinct from that of other
sources.  Selection criteria based on this signature can be applied to
nearby galaxies in which PNe appear as point sources.  We applied
these criteria to the whole area of \MA\ as scanned by the SDSS,
selecting 167 PN candidates that are located in the outer regions of
\MA.  The spectra of 80 selected candidates were then observed with
the 2.2m telescope at Calar Alto Observatory. These observations and
cross-checks with literature data show that our method has a selection
rate efficiency of about 90\%, but the efficiency is different for the
different groups of PNe candidates.

In the outer regions of \MA, PNe trace different well-known
morphological features like the Northern Spur, the NGC\,205 Loop, the
G1 Clump, etc.  In general, the distribution of PNe in the outer
region $8<R<20$~kpc along the minor axis shows the ``extended disk''
-- a rotationally supported low surface brightness structure with an
exponential scale length of $3.21\pm0.14$~kpc and a total mass of
$\sim10^{10} M_{\sun}$, which is equivalent to the mass of M\,33.  We
report the discovery of three PN candidates with projected locations
in the center of Andromeda\,NE, a very low surface brightness giant
stellar structure in the outer halo of \MA.  Two of the PNe were
spectroscopically confirmed as genuine PNe.  These two PNe are located
at projected distances along the major axis of $\sim$48~Kpc and $\sim$41~Kpc
from the
center of \MA\ and are the most distant PNe in \MA\ found up to now.

With the new PN data at hand we see the obvious kinematic connection
between the continuation of the Giant Stream and the Northern Spur. We
suggest that 20 -- 30\% of the stars in the Northern Spur area may
belong to the Giant Stream.  In our data we also see a possible
kinematic connection between the Giant Stream and PNe in Andromeda\,NE,
suggesting that Andromeda\,NE could be the core or remnant of the
Giant Stream.  Using PN data we estimate the total mass of the Giant
Stream progenitor to be $\approx10^9 M_{\sun}$.  About 90\% of its
stars appear to have been lost lost during the interaction with \MA.
\end{abstract}


\section{Introduction}

Planetary nebulae (PNe) arise from intermediate- and low-mass stars,
which makes them an excellent probe of the dynamics of these stars in
nearby galaxies. Their bright emission lines can provide accurate
line-of-sight velocities with a minimum of telescope time.  Therefore
spectroscopy of PNe can be used as a powerful tool for the study of
the kinematics of nearby galaxies \citep[e.g.,][ hereafter HK04,
PFF04, and M06]{HK04,PFF04,Mer06} and for the mapping of stellar
streams around massive galaxies such as \MA\ \citep{Mer03,Mor03}.  The
spectra of individual PNe in nearby galaxies also provide chemical
abundances of certain elements in their progenitor stars
\citep[e.g.,][]{W97,Richer98,Fornax,SALT_Sgr,Sa09,Magr09a,Kwitter12,Sanders12}.
They complement photometric or spectroscopic metallicity information
traditionally derived from old red giants or from H\,{\sc ii} regions
by providing a probe of the abundances of intermediate-age
populations.  

Searches for PNe are usually based on narrow-band imaging in the
H$\alpha$ and [O\,{\sc iii}] $\lambda$5007 bandpasses in which PNe can
emit 15--20\% of their total luminosity
\citep[e.g.,][]{Magr03,Magr05a}. Occasionally special instruments like
the Planetary Nebula Spectrograph \citep[PM.S;][]{Douglas02} are
employed for the simultaneous identification of PNe and the
measurement of their radial velocities \citep[e.g.,][]{Mer06}, or
integral field spectrographs are used in small fields of view with
particularly high crowding in order to identify and measure [O\,{\sc
iii}] $\lambda$5007 emission against a pronounced stellar background
\citep{Pastorello13}.  Other searches, especially photometric ones,
aim at covering as large an area as possible.

The Sloan Digital Sky Survey (SDSS) \citep{york00,stou02} was an
imaging and spectroscopic survey that covered about one quarter of the
celestial sphere.  The imaging data were collected in drift-scan mode
in the five bandpasses $u, \ g, \ r, \ i$, and $z$
\citep{fuku96,gunn98,hogg01}.  The images were subsequently processed
with special data reduction pipelines to measure the photometric,
astrometric, and structural properties of all detectable sources
\citep{lupt02, stou02, smit02, pier03} and to identify targets for
spectroscopy. The SDSS passbands were carefully chosen to provide a
wide color baseline, to avoid night sky lines and atmospheric OH
bands, to match passbands of photographic surveys, and to guarantee
good transformability to existing extragalactic studies.

The SDSS has been used extensively for the detection and
characterization of objects with special characteristics, including
different types of emission-line objects
\citep[e.g.,][]{Rich02,Kni03,Kni04,PT11,Tanaka12,Zhao13}.
Since the detected flux from PNe comes almost entirely from nebular
emission lines in the optical, the range of colors characteristic of
PNe is defined by the ratios of these emission lines and their
corresponding contributions in different SDSS passbands. Some of these
colors may be expected to be similar to the colors of emission-line
galaxies \citep[ELGs; see, e.g.,][]{Kni04} and should be usable for
the detection of PNe based on SDSS photometry.

In this paper, we present a method designed to detect PN candidates
based on their SDSS colors.  Using a sample of known PNe, we isolate
a region in the SDSS $ugri$ two-color diagram in which the probability
of an object to be a PN is very high.  In \S\ref{txt:method} we
describe the detection method. We apply this method to the \MA\ region
scanned by the SDSS in 2002 \citep[DR6,][]{DR6}.  In
\S\ref{txt:obs_red} the follow-up observations and data reduction are
described.  In \S\ref{txt:compar} we compare our data with other
surveys for PNe in \MA.  In \S\ref{txt:efficiency} the resulting
detection efficiency of the method is discussed.  In
\S\ref{txt:M31PNe} we present our results for the new PNe in \MA\ and
discuss them.  A summary is presented in \S\ref{txt:summary}.
%For the remainder of this paper, we adopt the revised solar value of
%the oxygen abundance, 12+$\log$(O/H) = 8.66 \citep{Asp04}.
For the remainder of this paper, we assume a distance to \MA\ of
760\,kpc \citep{vand99}.

%---------------------------------------------------------------
\section{The Method}
\label{txt:method}
%---------------------------------------------------------------

%---------------------------------------------------------------
\subsection{Primary Selection Criteria}
\label{txt:Sel_determ}
%---------------------------------------------------------------

In order to develop a method for the selection of PN candidates from
Sloan Digital Sky Survey (SDSS) imaging data based on photometric
criteria, we used an SDSS scan of \MA\ reduced with the standard SDSS
photometric pipeline \citep[see][]{Zucker04a}.  Using 37 PNe in \MA\
from \citet[][NF87 hereafter]{NF87} and one PN from \citet{JF86} we
constructed a test sample of previously known PNe in this region.  We
re-identified these PNe in the SDSS data and developed selection
criteria on the basis of their SDSS colors and magnitudes.  Because the
standard SDSS pipeline (PHOTO) does not work properly in very crowded
fields, there are {\em no}\ SDSS data in the central area of \MA\
where many of the previously identified PNe in \MA\ are located.  In
addition, eight PNe from the list of 37 from \citet{NF87} could not be
recovered since they are either located close to diffraction spikes of
bright stars on SDSS images or lie in regions where the SDSS source
detections are incomplete due to crowding.  Our final test sample
contained 30 PNe.

\begin{figure*}[th]
\begin{center}
\includegraphics[width=18.0cm,angle=0,clip=true]{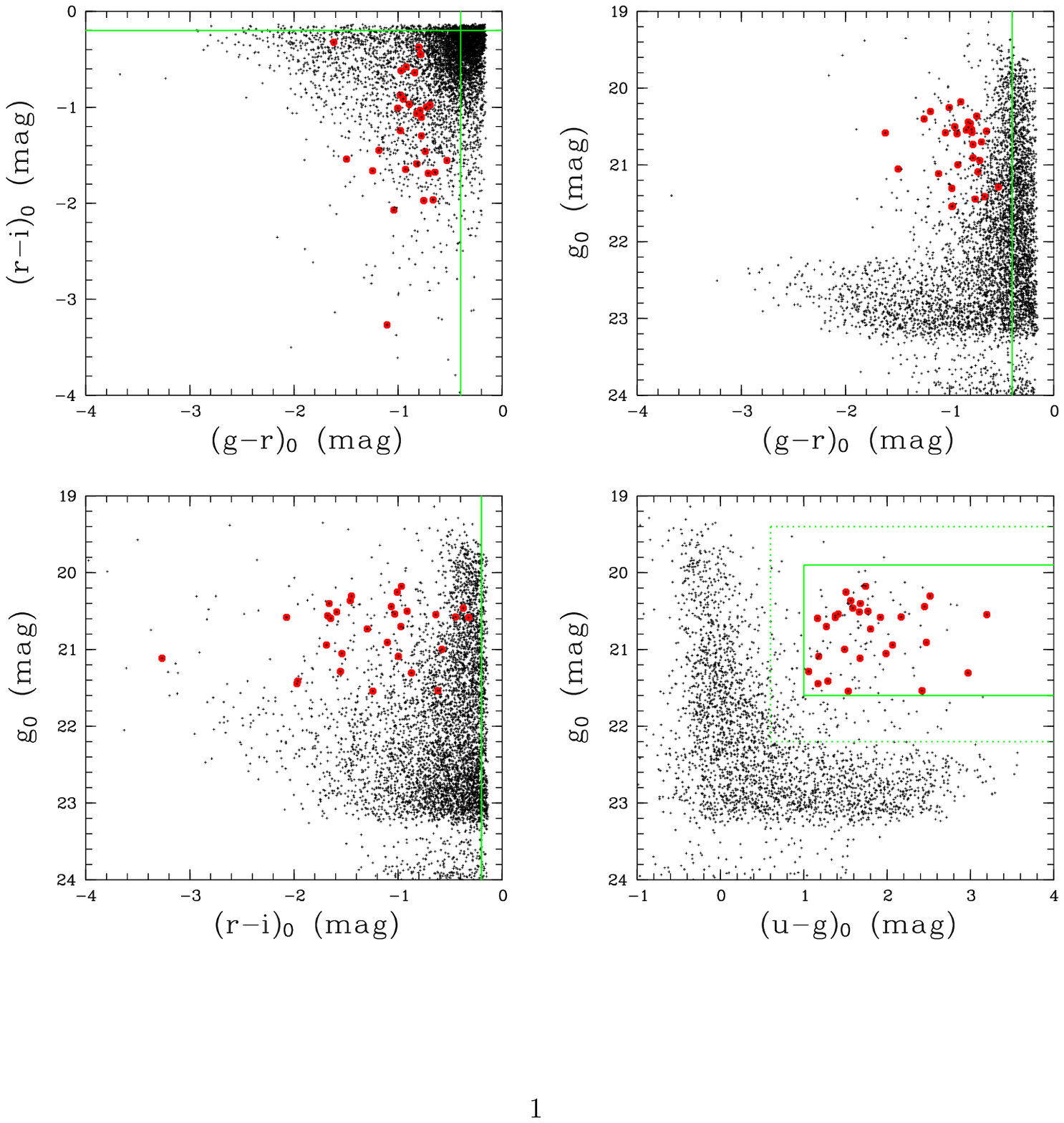}
\end{center}
\caption{\label{fig:Sel_crit}
Color-magnitude diagrams for sources from SDSS \MA\ data to which we
applied our selection procedure.  All previously known re-identified
PNe from \citet{NF87} and \citet{JF86} are shown with red filled
circles. Their locus defines our ``first priority'' candidates.  Our
color-magnitude criteria for selecting these candidates are shown by
solid lines.  Additional dotted lines in the bottom-right $g_0$ vs.\
$(u-g)_0$ diagram show softer criteria for the selection of the
``second priority'' candidates.  Within these lines only sources
remaining after visual verification are plotted (see
Section~\ref{txt:Sel_determ} for more details).
}
\end{figure*}

The location of the known PNe from our test sample and of other
stellar sources from the SDSS \MA\ data in different color-magnitude
diagrams (CMDs) is shown in Figure~\ref{fig:Sel_crit}.  Throughout
this paper, we use magnitudes resulting from point spread function
(PSF) fitting in the SDSS photometry pipeline, as these magnitudes
give the best results for point sources in the SDSS.  To minimize any
reddening effects, the measured magnitudes for each object and each
filter were corrected using the extinction values from the
\citet{Schl98} maps prior to further analysis.

One of the basic characteristics for PNe is the magnitude m$_{5007}$,
which is an equivalent of the V-magnitude calculated from the flux of
the emission line [O\,{\sc iii}] $\lambda$5007 \citep{J89}.  The SDSS
filter $g$ (central wavelength 4686 \AA) covers this emission line in
a region very close to the maximum of its response. Since PNe at the
distance of \MA\ are objects without detected continuum we assume
that the $g$ magnitudes for these objects have to be very close to the
m$_{5007}$ magnitudes (see Section~\ref{txt:compar} for the final
comparison). Our final criteria to recover all objects of the test
sample from the SDSS photometric database are:
\begin{eqnarray}
\label{eqn:final_select}
	     {\rm  Object~type} &   = &  {\rm star}   \nonumber   \\
	     {\rm  Magnitude~type} & = &  {\rm PSF}    \nonumber   \\
	     19\fm9 \le  g_0 & \le &  21\fm6 \nonumber \\
		   (g - r)_0 & \le & -0\fm4            \\
		   (r - i)_0 & \le & -0\fm2  \nonumber \\
		   (u - g)_0 & \ge &  1\fm0  \nonumber
\end{eqnarray}

As Figure~\ref{fig:Sel_crit} demonstrates the most important CMD is
$(u - g)_0$ versus $g_0$, in which the locus of all PNe from our test
sample is clearly separated from the location of most other stars in
this diagram.  The very red $(u - g)_0$ colors of PNe are defined by
the existence of strong [O\,{\sc iii}] $\lambda$5007 emission in the
$g$ filter and the absence of any strong emission lines in the $u$
filter. The limiting magnitude for the bright end of the PN
distribution, $g_0 = 19\fm9$, was selected on the basis of the
distance modulus $\rm (m - M)_0 = 24.4$ to \MA\ \citep[][]{vand99} and
the absolute magnitude cut-off of the PN [O\,{\sc iii}] $\lambda$5007
luminosity function (PNLF) in massive galaxies with a large population
of PNe \citep[$M_{5007} = -4\fm47$;][]{Ciar02}.  The limiting
magnitude for the faint end and the bluest limiting $(u - g)_0$ color
of the PNe was selected such as to recover all objects of the test
sample.

The locus of the PNe from the test sample defines the selection
parameter range for our ``first priority'' candidates.  Since these are
objects whose photometric properties match those of the known PNe from
the test sample, they are considered to be very likely PNe
as well. However, it is conceivable that ``true'' PNe are located not
only in this region of parameters, but have some spread around it.
PNe could have brighter $g_0$ magnitudes because $g$ is not exactly
identical with a narrow-band m$_{5007}$ filter. We also expect PNe
with fainter $g_0$ magnitudes, because the PNLF for \MA\ has been traced down
to 6 mag fainter from its bright cut-off \citep[][]{Mer06}.  PNe could
also have redder $(u - g)_0$ colors in the case that the [O\,{\sc
iii}] $\lambda$5007 line is not as strong as for objects from the test
sample.  

For these reasons, based only on the distribution plotted in the $(u -
g)_0$ vs.\ $g_0$ CMD, we additionally defined more relaxed selection
criteria of $19\fm4 \le g_0 \le 22\fm2$ and $(u - g)_0 \ge 0\fm6$ to
identify ``second priority'' candidates. We realized that we can
assess how good or bad these relaxed criteria are only after getting
additional information or confirmation observations for the selected
candidates.  These softer criteria for the selection of the second
priority candidates are indicated with dotted lines in the
bottom-right $(u-g)_0$ vs.\ $g_0$ diagram  in
Figure~\ref{fig:Sel_crit}.  As can be seen from this figure, for
magnitudes fainter than $g_0 = 22\fm2$, the CMD data become very
uncertain because of the incompleteness of the SDSS \MA\ data
themselves.  The completeness of the SDSS \MA\ data varies from field
to field because of, for instance, variable seeing during the
observations and photometry pipeline problems caused by crowding.
Because of the high stellar density in the \MA\ region, the
incompleteness is higher than in standard SDSS imaging data,
which have 95\% completeness for point sources at the level
of $r \sim 22\fm2$ \citep{DR2}.

We then applied our criteria to the whole area of \MA\ observed by the
SDSS in 2002 \citep[][]{DR6} in order to select PN candidates.  Since
the star-galaxy separation in the SDSS is better than 90\% at $r =
21\fm6$ \citep{DR1}, but worsens for fainter magnitudes, fainter PNe
may have been wrongly identified as (slightly) extended sources in the
SDSS database.  In our case $r = 21\fm6$ can be transformed to $g_0
\sim 20\fm6$ using $(g-r)_0 \sim -1$ from Figure~\ref{fig:Sel_crit},
and could be even brighter.  In order not to lose such potential PNe,
we applied the same color and magnitude selection criteria as listed
above to extended objects and selected additional PN candidates of
first and second priority. Subsequently all candidates were visually
inspected to eliminate recognizably false detections, such as
diffraction spikes of bright stars and clearly extended objects.  

We also visually checked our selected candidates using the images of
the ``Survey of Local Group Galaxies Currently Forming Stars''
\citep[SLGG hereafter;][]{Mas06,Mas07}.  In this survey, imaging was
obtained in broadband {\em UBVRI} filters and in narrow-band filters
centered on the H$\alpha$ and [O\,{\sc iii}]$\lambda$5007 lines. Using
the SLGG we marked those of our candidates with obvious emission in
the [O\,{\sc iii}] $\lambda$5007 and H$\alpha$ narrow-band filters and
removed all candidates without such emission.  This work was done only
for that part of our sample covered by the SLGG images.

\begin{figure}[h]
\begin{center}
\includegraphics[width=10.0cm,angle=-90,clip=true]{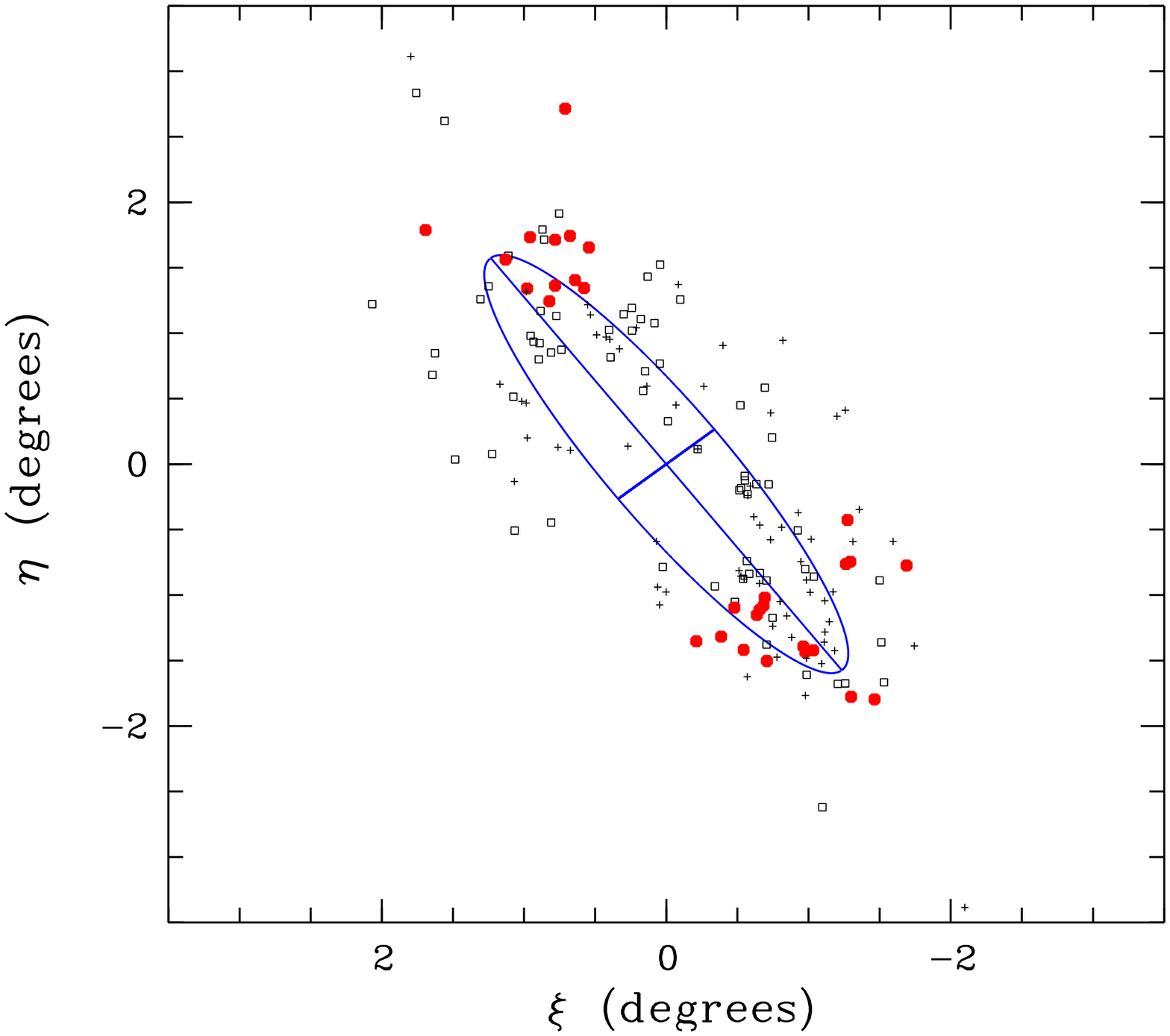}
\end{center}
\caption{\label{fig:Sel_pos}
Positions of selected candidates relative to the center and
orientation of \MA.  The ellipse has a semimajor axis of 2~degrees
($\approx27$~kpc) and represents an inclined disk with $i=77.5$. The
optical disk of \MA\ lies well within this boundary \citep{Fer02}.
First priority candidates are marked with squares and second priority
candidates with crosses.  PNe from \citet{NF87} and \citet{JF86} are
shown with filled (red) squares.  Three PN candidates are located in
the area of Andromeda NE \citep{Zucker04a} in the range of the
coordinates (2.0--2.5, 2.5--3.0).
}
\end{figure}

In the end, we detected a total of 167 PN candidates, 100 first
priority and 67 second priority.  By design, all PNe from the
test sample were detected as candidates of the first priority.  We
list all selected \MA\ PN candidates in the SDSS data in
Table~\ref{tbl-1}.  This table contains the designations (column 1),
coordinates (columns 2--3), PSF magnitudes corrected for the
extinction and magnitude uncertainties as given by the SDSS (columns
4--8), and the priority type (column 9).  The table also shows the
result of our spectroscopic follow-up observations and visual checks
with data from the SLGG (column 10; see Section \ref{txt:obs} for
detailed explanations), cross-identifications with other catalogs of
PNe in \MA\ (column 12), and the possible association of objects from
our sample with some structures of the outer region of \MA\ (column
11).  In Figure~\ref{fig:Sel_pos} we show the positions of all
selected candidates of both priorities in our survey using a standard
coordinate projection centered on \MA.

%---------------------------------------------------------------
\subsection{Possible Contaminants}
\label{txt:contam}
%---------------------------------------------------------------

In order to understand possible contamination by point sources such as
QSOs and stars, we analyzed the distribution of our candidates as compared to the
known loci of quasars of different redshifts as shown and discussed in
\citet{Rich02} and \citet{QSO10}.  Our analysis shows that our
criteria select star-like sources that are far from both the QSO loci
and the loci of the Galactic stars fainter than $g_0 \sim 15.3$ mag.
Only a few data points from stars are located in our areas of
interest, which would yield perhaps 1--10 objects per 1000 deg$^2$.
Since the SDSS \MA\ data studied in this paper cover only
about 26 deg$^2$ in total \citep{DR6},
including the central part of \MA, such contamination should thus be
extremely small.

\begin{figure}[h]
\begin{center}
\includegraphics[angle=-90,width=10cm,clip=true]{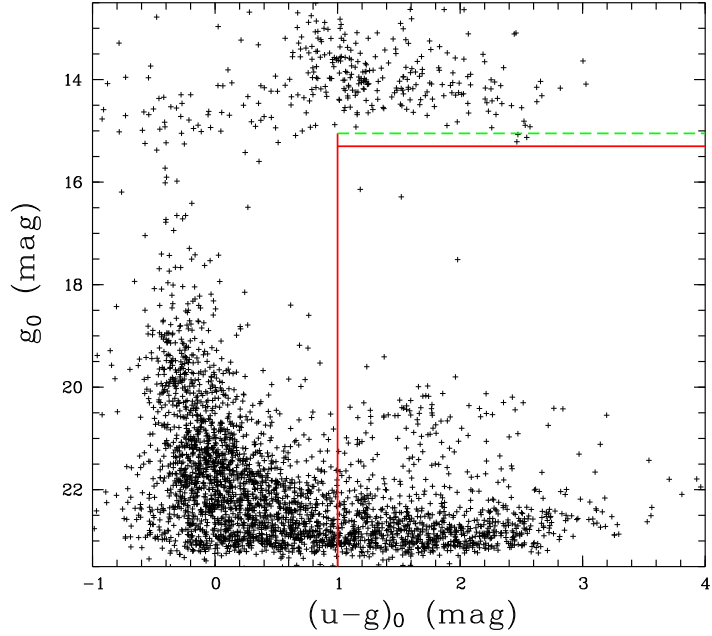}
\caption{\label{fig:All_stars}
Color-magnitude diagram for all stellar sources from SDSS \MA\ data
selected with the criterion $(g - r)_0 \le -0\fm4~~\&~~(r - i)_0 \le
-0\fm2$.  Our color-magnitude criterion $(u - g)_0 \ge 1\fm0$ for the
selection of the first priority candidates is shown with a vertical
line.  Foreground stars from the Galaxy are located in the region
brighter than $g_0=$15\fm3 that is shown with a horizontal solid line.
The magnitude $g_0=$15\fm05, which corresponds to a distance of
$\sim80$~kpc for an absolute magnitude cutoff of the PNLF, $M_{5007} =
-4\fm47$, is shown with a horizontal dashed line.  Three star-like
sources with $g$-band magnitudes between 16\fm0 and 19\fm0 and $(u-g)
> 1\fm0$ are false detections in the areas around bright saturated
stars.
}
\end{center}
\end{figure}

For stars brighter than $g_0 \sim 15.3$ mag foreground stars from the
Galaxy begin to contribute as shown in Figure~\ref{fig:All_stars},
where the CMD for all available stellar sources from the \MA\ SDSS
data is plotted after applying the criteria $(g - r)_0 \le
-0\fm4~~and~~(r - i)_0 \le -0\fm2$.  Figure~\ref{fig:All_stars} shows
that $g_0 \sim 15.3$ mag is a natural limit of our color-selection
method and is far away from the limiting magnitude of 19\fm4 we
used for the \MA\ data.

We also estimated the amount of possible contamination by background
ELGs using data from \citet{Kni04} for galaxies with strong emission
lines from SDSS Data Release 1 \citep{DR1}.  With our color criteria,
only two of the ELGs from the SDSS catalog of \citet{Kni04} would be
selected.  These two have very high ([O\,{\sc iii}]
$\lambda$5007/H$\beta$) line ratios of 6.5 and 7.2, respectively,
which are very close to the characteristics of the PNe from the test
sample (see Section~\ref{txt:efficiency}).

\begin{figure}[h]
\begin{center}
\includegraphics[angle=-90,width=10.5cm,clip=true]{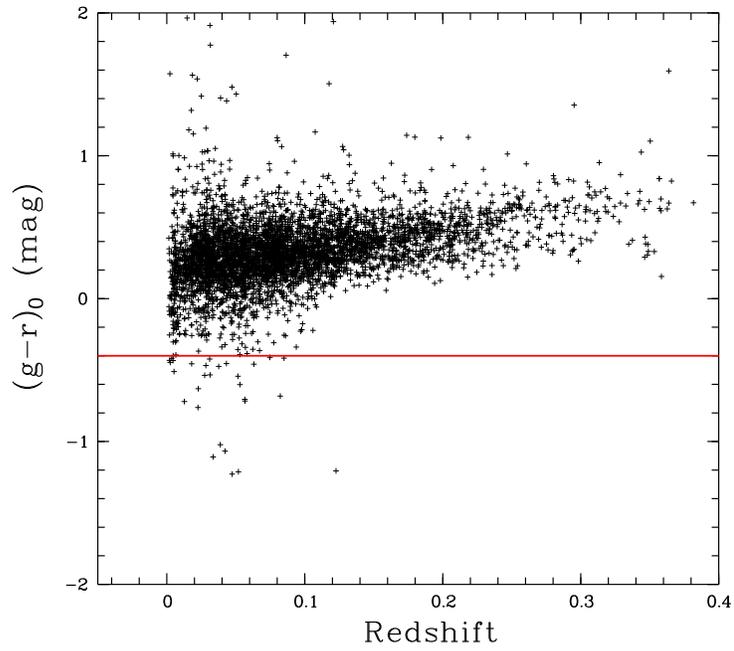}
\caption{\label{fig:ELGs_red}
The distributions of $(g-r)_0$ colors from PSF photometry for ELGs
from \citet{Kni05} versus redshift.  Our selection criterion for the
$(g-r)_0$ color is shown with a solid red line.
}
\end{center}
\end{figure}

Considering that the above ELG catalog is based on the SDSS Data
Release 1, which covers an area of 1360 deg$^2$, the number of
possible contaminants in our \MA\ region is obviously very small.
Furthermore, the two recovered ELGs are clearly very extended objects
and would therefore certainly have been rejected during the visual
inspection. These two ELGs are both relatively bright sources with
total magnitudes $r \le 17\fm77$ (this was the spectroscopic target
magnitude limit for galaxies in the SDSS-I). In contrast, it is
conceivable that very distant, very faint ELGs might appear as
star-like sources instead.  As can be seen in
Figure~\ref{fig:ELGs_red}, the $(g-r)_0$ color for ELGs increases with
redshift.   Beyond a redshift of 0.1 all ELGs will lie outside of our
color selection area.  Hence we may assume that contamination of our
PN sample by faint ELGs is negligible.

We conclude that our color-selection method for PN candidates using
SDSS $ugri$ filters appears to work very well for point-like sources
at the distance of \MA\ with a very low probability for the selected
sources to be contaminated with other types of objects.  Potentially,
this method can work up to the bright limit of $g_0 \sim 15.3$, which
corresponds to a distance of $\sim 90$~kpc (for the absolute magnitude
cutoff of the PNLF of $M_{5007} = -4\fm47$).
For extended sources at the distance of \MA, the probability that the
selected candidates could be contaminated by nearby (redshift $\le$
0.1) ELGs with strong emission lines is also very low.  But this
contamination will surely grow when the $(u-g)_0$ color criterion is
relaxed (see Section~\ref{txt:efficiency} for more details).

%---------------------------------------------------------------
\section{Spectroscopic Follow-up Observations}
\label{txt:obs_red}
%---------------------------------------------------------------

%---------------------------------------------------------------
\subsection{Observations}
\label{txt:obs}
%---------------------------------------------------------------

Spectroscopic follow-up observations of a subset of our PN candidates in
\MA\ were carried out in 2004 October 7 to 14 at Calar Alto
Observatory with the Calar Alto Faint Object Spectrograph (CAFOS) at
the 2.2\,m telescope.  During the eight nights of observations in
service mode a total of 80 PN candidates of both first and second
priorities were observed under variable weather conditions. The seeing
varied from 0\farcs8 to 2\farcs5.  A long slit whose width was
adjusted depending on the seeing (1\arcsec\ -- 2.5\arcsec) was used in
combination with a G-100 grism (87\,\AA\,mm$^{-1}$, first order).  The
spatial scale along the slit was $0\farcs53$~pixel$^{-1}$.  The
detector was a SITE 2K$\times$2K CCD, which we used without binning.
The resulting wavelength coverage was $\lambda$\,4200 --
$\lambda$\,6800\,\AA\ with maximum sensitivity at $\sim$~6000\,\AA.
The obtained dispersion was $\sim$1.9--2\,\AA/pixel, leading to a
spectral resolution of $\sim$~4 -- 6\,\AA\ (FWHM).  The exposure times
were adjusted according to the target brightness and weather
conditions and ranged from 15 to 30 minutes per target.  In addition,
the flux standard star Hiltner~102 was observed at least once per
night, and Hg--Cd reference spectra for wavelength calibration were
obtained, complemented by the usual dome flatfields, bias, and dark
exposures.  

We have marked all spectroscopically observed PNe in Table~\ref{tbl-1}.
Column 10 specifies which of the PNe were observed at Calar Alto
Observatory during our spectroscopic follow-up campaign.  Confirmed
PNe are marked with the flag value ``1''. Three candidates that were
not detected as PNe in our follow-up but that were confirmed later
via cross-identifications with other catalogues as real PNe are marked
with the flag value ``2''. PN candidates that were not found in our
follow-up observations are marked with the flag value ``3''.  We did
not reject these unconfirmed candidates because all of them are
fainter than g=21\fm5 and possibly were not identified correctly with
the 2.2\,m telescope under poor weather conditions.

%---------------------------------------------------------------
\subsection{Data Reduction}
\label{txt:red}
%---------------------------------------------------------------

\begin{figure}[th]
\begin{center}
\includegraphics[width=12.0cm,angle=-0,clip=true]{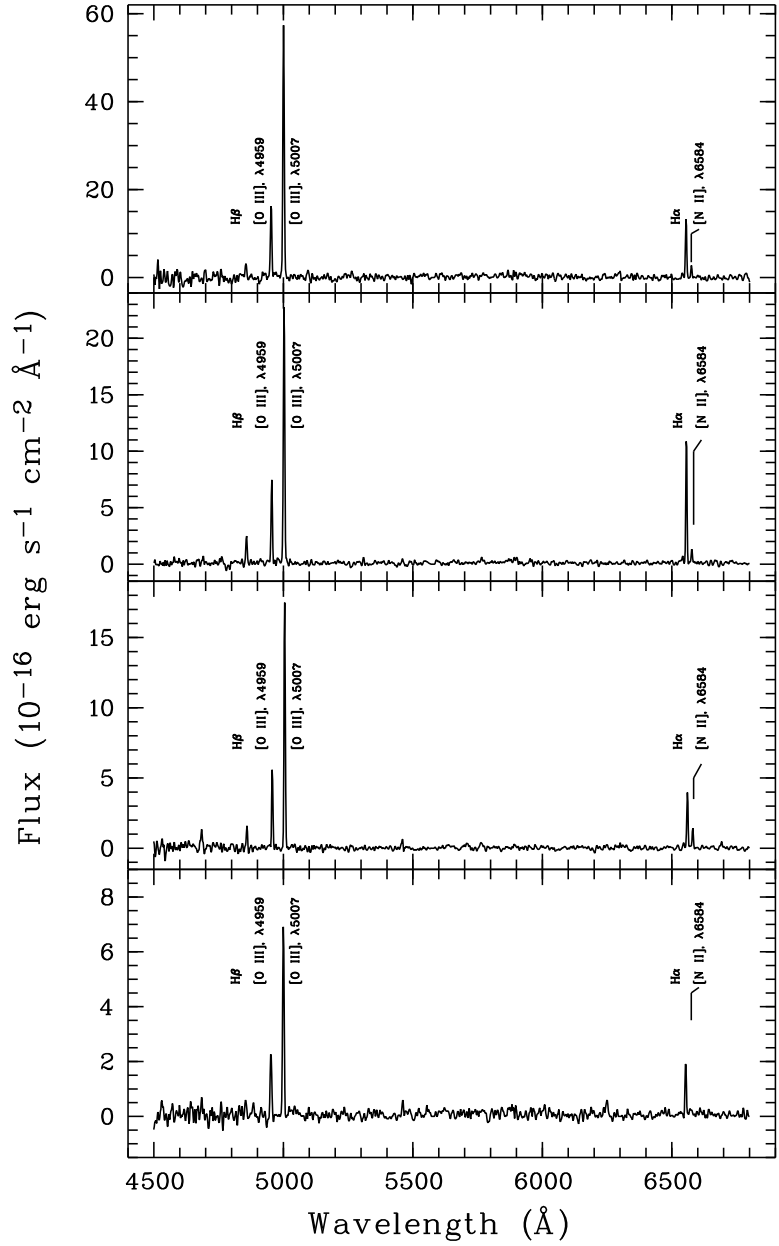}
\caption{\label{fig:SDSS_PN_spec}
Four typical spectra for our sample with different signal-to-noise
ratios and $g$ magnitudes ranging from $\sim$20 to 21\fm6.  All
spectra show the strongest emission lines [O{\sc iii}]
$\lambda$4959,5007 and H$\alpha$.
}
\end{center}
\end{figure}

The two-dimensional spectra were bias-subtracted and flat-field
corrected using IRAF\footnote{IRAF: the Image Reduction and Analysis
Facility is distributed by the National Optical Astronomy Observatory,
which is operated by the Association of Universities for Research in
Astronomy, Inc. (AURA) under cooperative agreement with the National
Science Foundation (NSF).}.  Cosmic ray removal was done with the
FILTER/COSMIC task in MIDAS\footnote{MIDAS is an acronym for the
European Southern Observatory data reduction package -- Munich Image
Data Analysis System.}. We used the IRAF software routines IDENTIFY,
REIDENTIFY, FITCOORD, and TRANSFORM to perform the wavelength
calibration and to correct each frame for distortion and tilt.  The
accuracy of the velocity determination depends on careful wavelength
calibration of the spectra. The rms error in fitting the dispersion
curve was always less than 0.3~\AA, or 18~km~s$^{-1}$ at a wavelength
of 5000~\AA.  After flux calibration, one-dimensional (1D) spectra
were extracted from the reduced frames using the IRAF APALL routine to
allow us to measure the total flux.  The resulting reduced and
extracted spectra of typical observed PNe of different $g$ magnitudes
are shown in Figure~\ref{fig:SDSS_PN_spec}.

After 1D spectra were extracted, we used our standard method for
measuring emission-line intensities \citep{Kni04,Kni05}.  Briefly, our
programs determine the location of the continuum, perform a robust
noise estimation, and fit separate emission lines by a single Gaussian
superimposed on the continuum-subtracted spectrum.  The emission lines
H$\alpha$ $\lambda$6563 and [N\,{\sc ii}] $\lambda\lambda$6548,6584
were fitted simultaneously as a blend of three Gaussian features.  The
quoted uncertainties of the individual line intensities $\sigma_{\rm
tot}$ include two components: $\sigma_{\rm p}$ caused by the Poisson
statistics of line photon flux, and $\sigma_{\rm c}$, the uncertainty
resulting from the creation of the underlying continuum and calculated
using the Absolute Median Deviation (AMD) estimator.
%The total errors have been propagated to calculate the errors of all
%derived parameters.

Since our data are not of good quality, only the strongest emission
lines are detected in our spectra.  The emission lines [O\,{\sc iii}]
$\lambda$4959,5007 and H$\alpha$ are seen in all spectra.  The
emission lines H$\beta$ and [N\,{\sc ii}] $\lambda\lambda$6548,6584
are detected in most (but not all) spectra.  The average
signal-to-noise ratios for the detected lines are 33.3 for [O\,{\sc
iii}] $\lambda$5007, 13.2 for H$\alpha$ and 3.8 for H$\beta$.  Lines
with a signal-to-noise ratio less than one were rejected.  The strong
emission line He\,{\sc ii} $\lambda$4686 was also found in the
spectrum of SDSS~J005123+435321, which is located in the area of
Andromeda~NE \citep{Zucker04a} shown in Fig.~\ref{fig:SDSS_PN_spec}.
The observed flux of this line is about 65\% of the flux of the
H$\beta$ emission line.  

\begin{figure}[th]
\begin{center}
\includegraphics[width=9.0cm,angle=-90,clip=true]{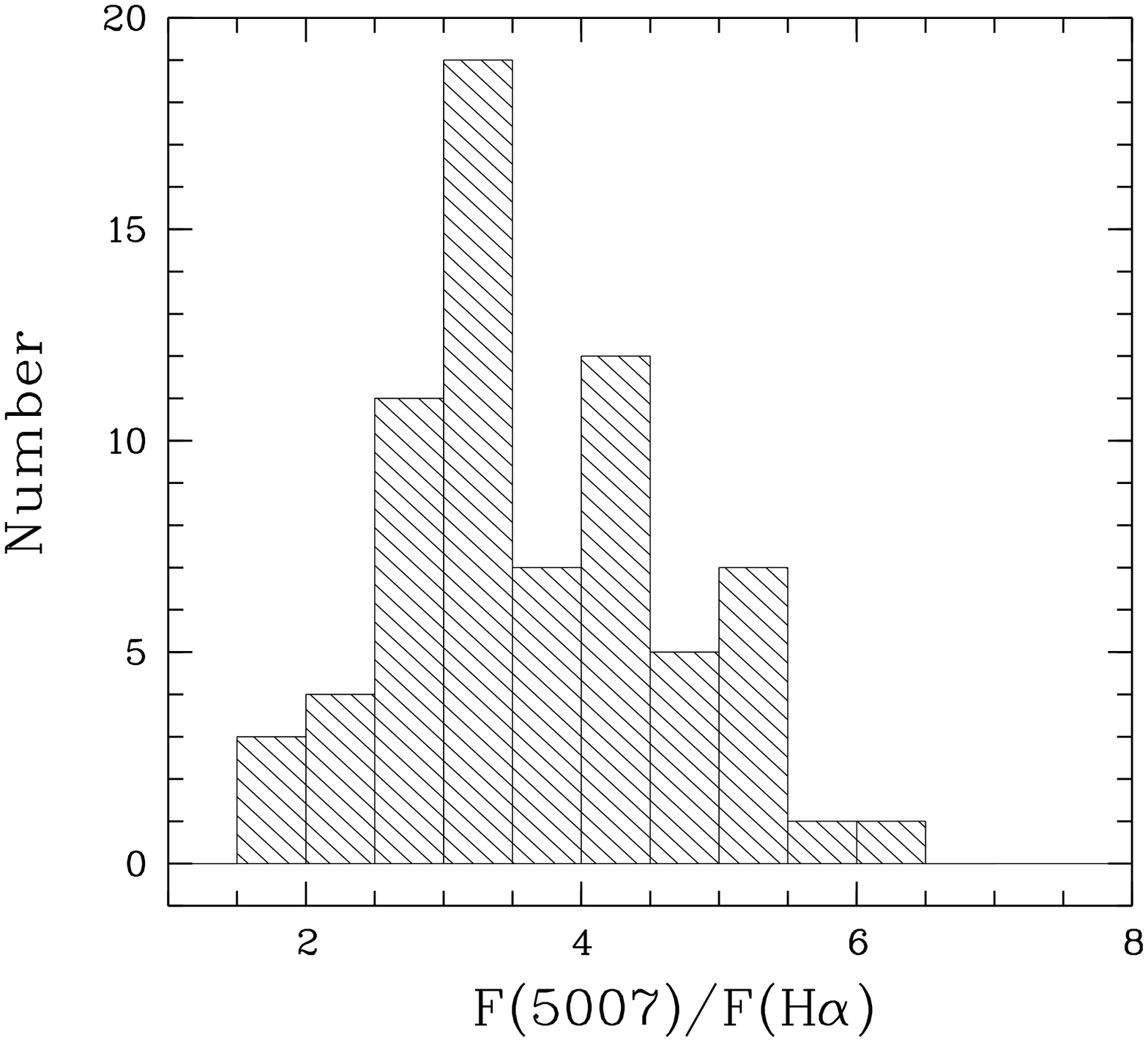}
\includegraphics[width=9.0cm,angle=-90,clip=true]{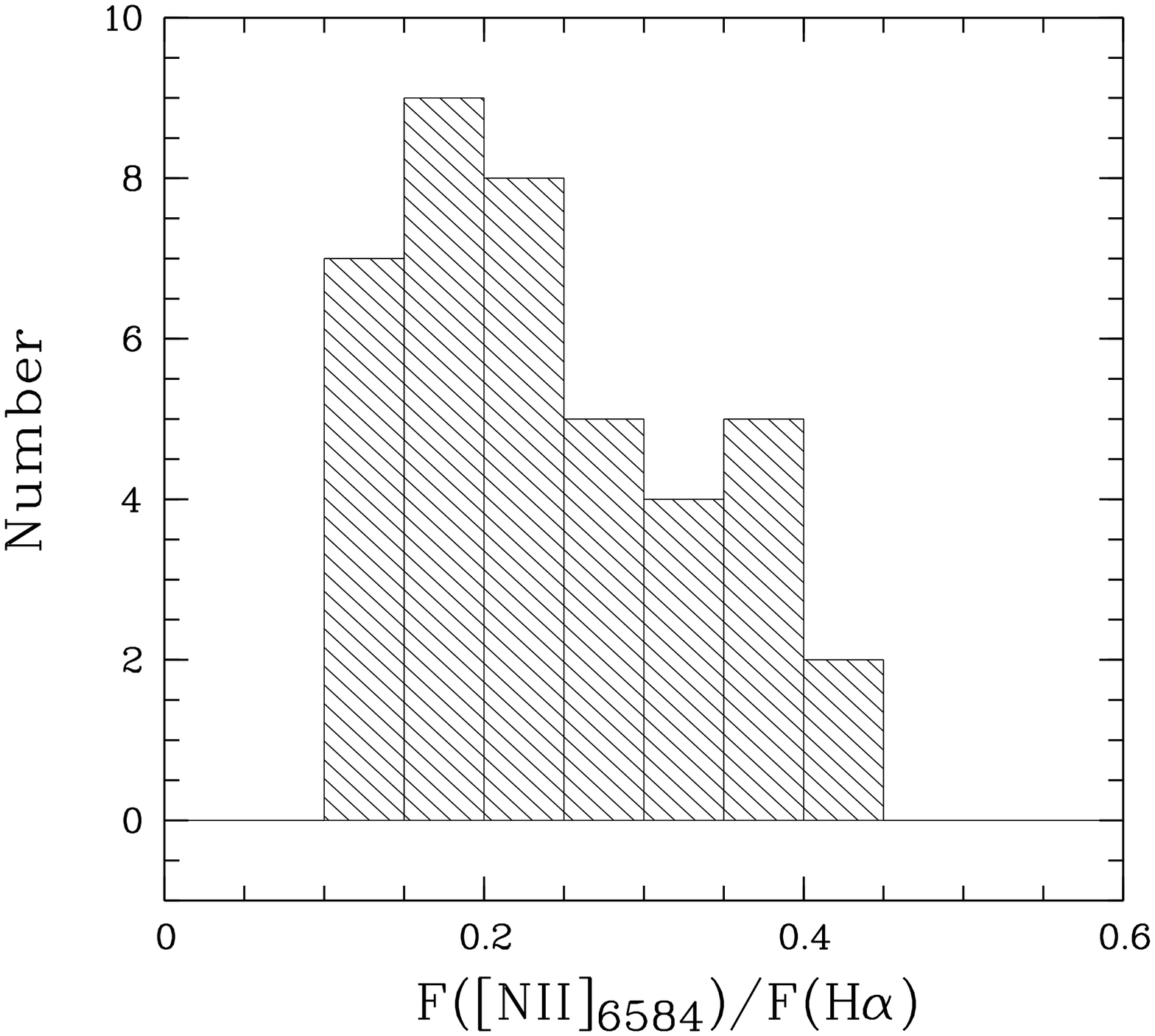}
\caption{\label{fig:lines_ratio}
Observed line ratios for PNe from our sample.
{\it Top}: The [O\,{\sc iii}] $\lambda$5007 to H$\alpha$ line ratio.
{\it Bottom}: The [N\,{\sc ii}] $\lambda$6584 to H$\alpha$ line ratio.
}
\end{center}
\end{figure}

\begin{figure}[th]
\begin{center}
\includegraphics[width=9.0cm,angle=-90,clip=true]{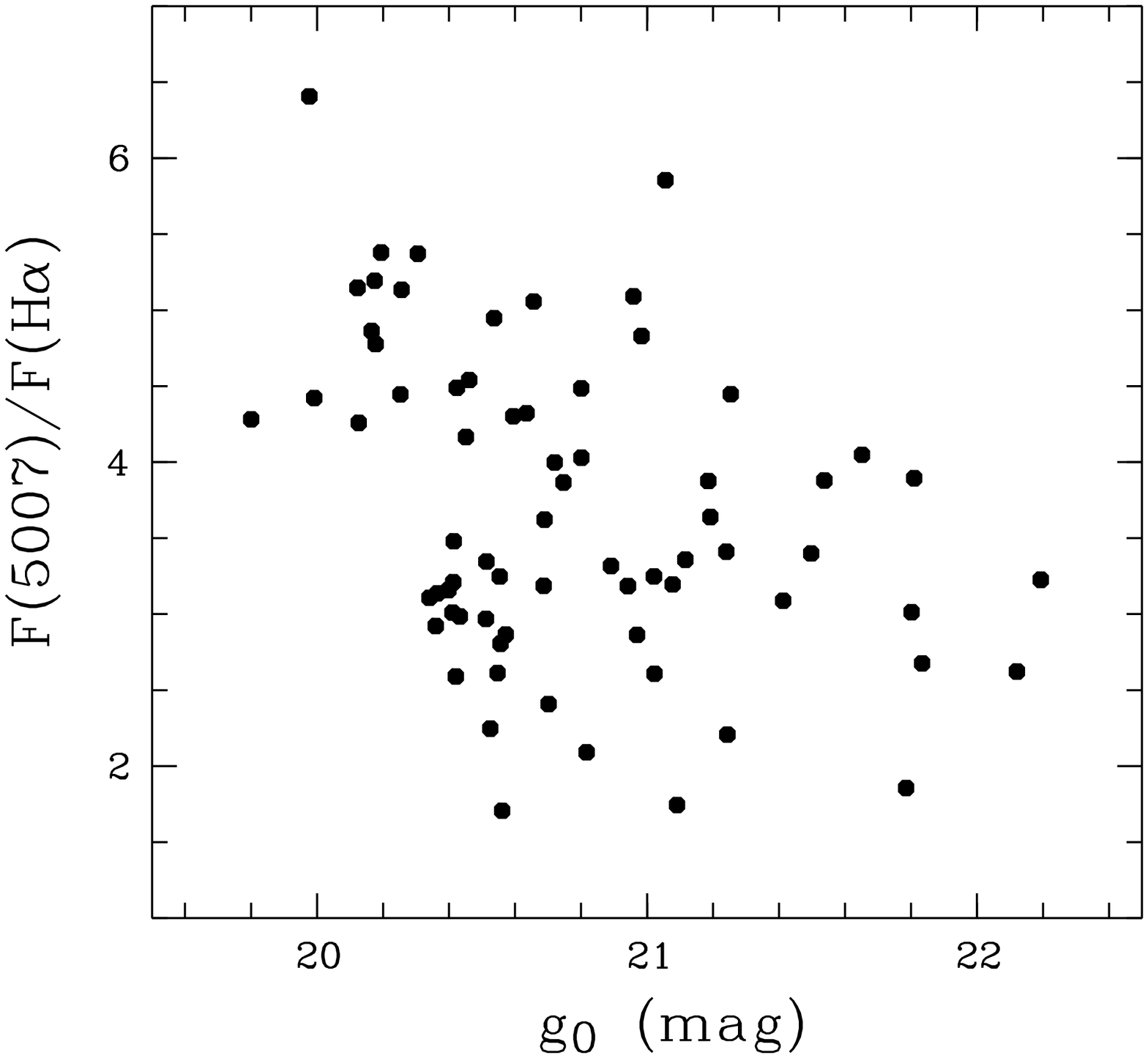}
\includegraphics[width=9.0cm,angle=-90,clip=true]{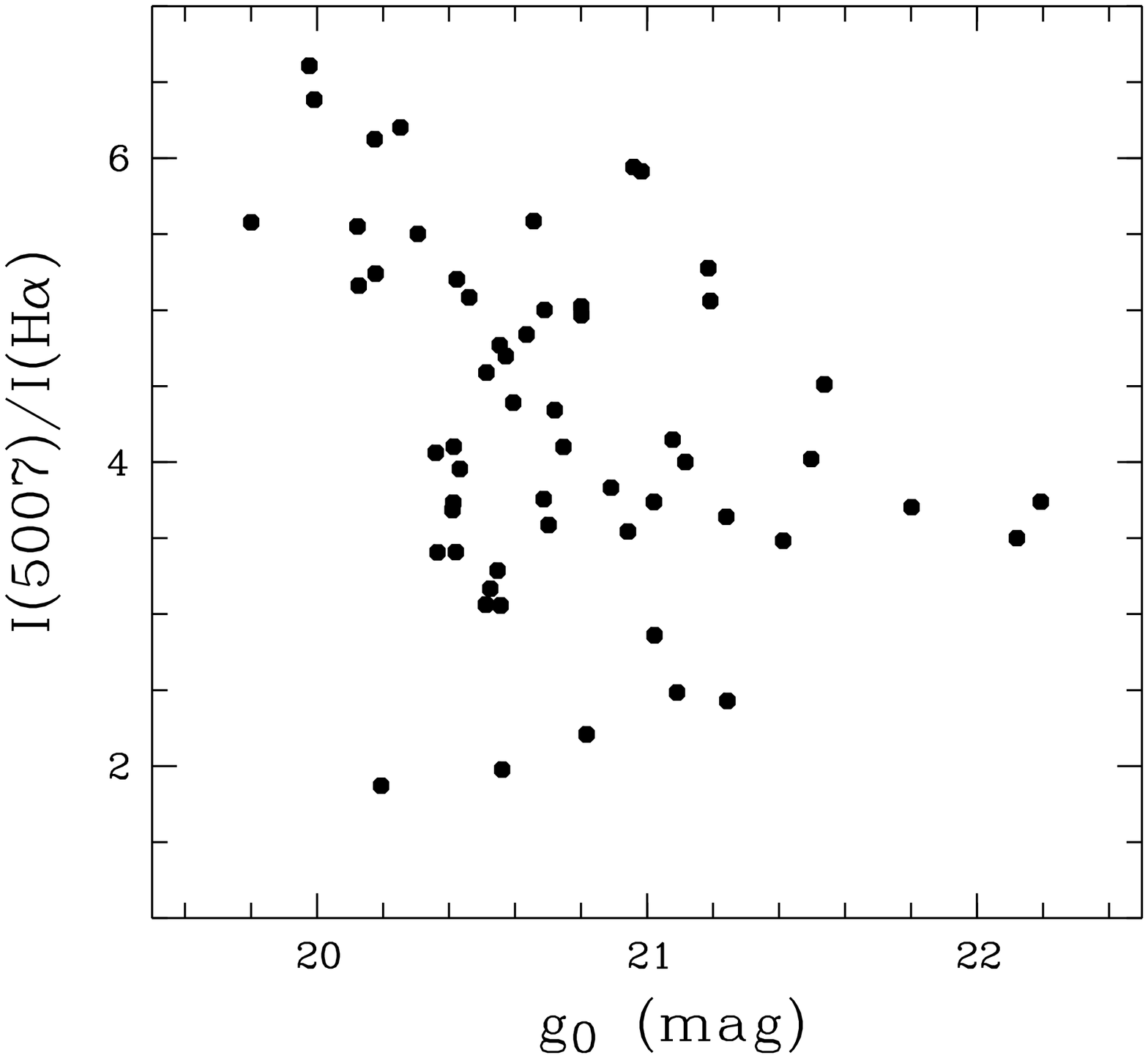}
\caption{\label{fig:mag_ratio}
{\it Top}: Observed [O\,{\sc iii}] $\lambda$5007 to H$\alpha$ line ratio versus
$g_0$ magnitude.
{\it Bottom}: [O\,{\sc iii}] $\lambda$5007 to H$\alpha$ line ratio versus
$g_0$ magnitude after correction for the calculated extinction.
}
\end{center}
\end{figure}

The quoted velocities were derived as mean values weighted by the
velocities determined from the individual lines.  The weights are
inversely proportional to the velocity accuracy for each line.  The
observed velocities were further corrected for the motion of the Earth
and transformed to heliocentric velocities.  The resulting radial
heliocentric velocities and their errors (column 2), the observed line
fluxes (columns 3 -- 6), and the derived extinction coefficient
$C$(H$\beta$) (column 7) based on the H$\alpha$/H$\beta$ ratio are
listed in Table~\ref{tbl-2}.  The distributions of the observed
[O\,{\sc iii}] $\lambda$5007~/~H$\alpha$ and [N\,{\sc ii}]
$\lambda\lambda$6548,6584~/~H$\alpha$ line ratios are shown in
Figure~\ref{fig:lines_ratio}.  The observed and the
extinction-corrected [O\,{\sc iii}] $\lambda$5007~/~H$\alpha$ line
ratios are plotted versus the $g_0$ magnitude in
Figure~\ref{fig:mag_ratio}.

It is worth noting here that in all cases of non-detection we see no
other emission lines in the spectra and do not see any continuum.  For
this reason we are not able to conclude anything about nature of the
candidates that are not identified as genuine PNe: no obvious ELGs,
QSOs, or identifiable special types of stars are detected.

%---------------------------------------------------------------
\section{Comparison with other data sets}
\label{txt:compar}
%---------------------------------------------------------------

After we constructed our sample and obtained follow-up observations,
data from four additional surveys for PNe in \MA\ were published.
These new data allow us to compare and to check for different
systematic effects or to test the external accuracy. The fifth
new survey of the center of \MA\ \citep{Pastorello13} is not included
here since the central regions are not resolved in our data.

\citet[][ hereafter HK04]{HK04} present positions and radial
velocities of a sample of 135 PNe, which were selected using
narrow-band imaging and follow-up spectroscopy in the area located to
the South and East of the nucleus of \MA.  \citet[][ hereafter
H06]{Hall06} published positions and velocities for 723 PNe located in
the disk and bulge of \MA. H06 used the conventional approach of
narrow-band imaging and fibre-fed spectroscopy.  \citet[][ hereafter
M06]{Mer06} present a catalogue of positions, magnitudes, and
velocities for 3300 emission-line objects (of which 2730 are probably
PNe) found by the Planetary Nebula Spectrograph in the area of \MA.
In our cross-identification work
we used a $2.5''$ search box, which is larger than the cited
astrometric accuracy of of both HK04 and M06.  

We did not compare our data with the data of H06, because the M06
sample includes 99\% of the H06 sample and the H06 sample is located
more in the central region of \MA.  We also do not compare with the
outer disk sample of \citet{Kwitter12}, since their 16 PNe were
selected from the M06 data.
Finally, we do not compare our data with the survey for PNe in globular
clusters in M31 \citep{Jacoby13} since due to crowding none of those
PNe and PN candidates are in our sample.

\begin{figure}[h]
\begin{center}
\includegraphics[width=15.0cm,angle=-90,clip=true]{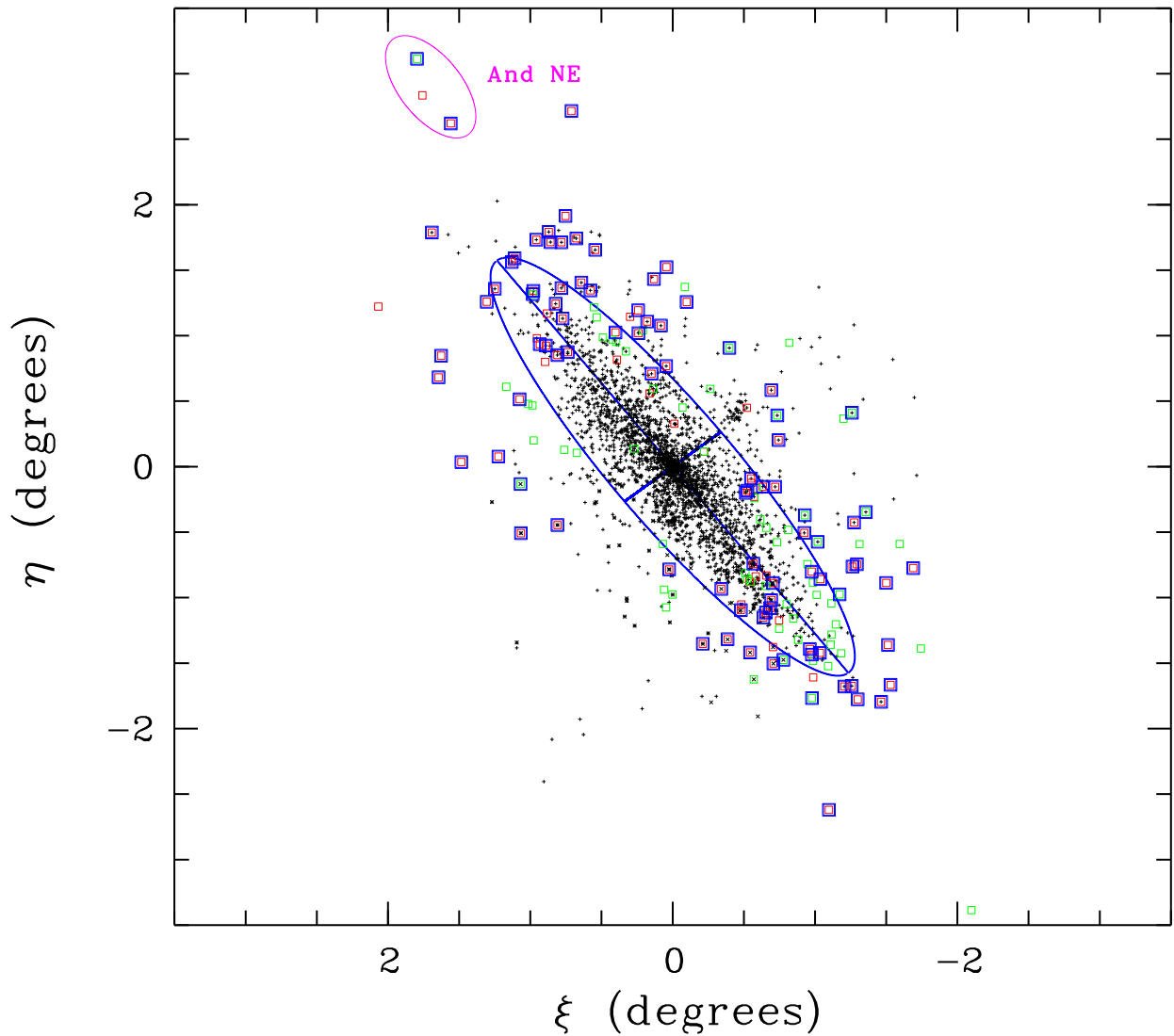}
\end{center}
\caption{\label{fig:Sel_pos1}
Positions of all PNe from the HK, H06, and M06 samples and selected
candidates from our work relative to the center and orientation of \MA.
The PNe from M06 and H06 are shown with plus signs (+).
The PNe from HK are shown with crosses (x).
First priority candidates from our sample are drawn with red squares
and second priority candidates are drawn with green squares.
All PNe from the test sample and all observed PNe from our sample
that were identified as real PNe are shown with larger blue squares.
Two of three PN candidates in the area of Andromeda NE were observed
and both of them were confirmed to be real PNe.
}
\end{figure}

Figure~\ref{fig:Sel_pos1} shows the spatial distribution of all PNe
from our sample, and of the samples of HK04, H06 and M06 relative to
the center and orientation of \MA.
The PN candidates from our
sample are shown with red (first priority) and green (second priority)
squares.  All observed PNe from our sample that are real PNe are
marked by blue squares that are larger in size than the other symbols.
All PNe from M06 and H06 are shown with plus signs (+).  All PNe from
HK are indicated by crosses (x).  In this figure, it is easy to see
the PNe that belong to both samples (as indicated by square symbols of
any color with a cross or plus inside) or that are new ones that were
discovered during this work (empty squares).

%---------------------------------------------------------------
\subsection{Velocities}
%---------------------------------------------------------------

As was described in Section~\ref{txt:Sel_determ} we used PNe from NF87
to define our selection criteria.  Therefore the PNe from the NF87
sample are also in our sample. We have nine PNe in common that were
re-observed at Calar Alto Observatory.  The weighted mean velocity
difference $\Delta v$(Our$-$NF87) is $14.3\pm6.6$ km~s$^{-1}$.  This
is very close to the systematic difference of 10.4 km~s$^{-1}$ that
HK04 found comparing their velocities with those of NF87.

In our final sample we have 17 PNe in common with HK04 (six of them
are from NF87), but only eight PNe that were re-observed at Calar
Alto.  The weighted mean velocity difference $\Delta v$(Our$-$HK04) is
$3.7\pm3.7$~km~s$^{-1}$, which means that we do not have any
substantial systematic offset in the velocities of our and of the HK04
data.  We have 12 PNe in common with H06, seven of which were
re-observed at Calar Alto.  Similarly, the weighted mean difference
$\Delta v$(Our$-$H06) is $1.2\pm2.4$~km~s$^{-1}$.  All these PNe from
H06 in our sample are also found by M06.

We have 66 PNe in common with M06 (many of them are in common with
NF87 and/or HK04 as well) of which 43 were re-observed at Calar Alto.
The weighted mean difference $\Delta v$(Our$-$M06) is $1.6\pm3.0$
km~s$^{-1}$ with a combined dispersion of 19.5~km~s$^{-1}$, implying
that we do not have any systematic offset between our and the M06
data. M06 found that the PN.S data have an uncertainty of
14~km~s$^{-1}$, thus our data have about the same 14~km~s$^{-1}$
uncertainty.

%---------------------------------------------------------------
\subsection{Astrometry and $g$ versus m$_{5007}$}
%---------------------------------------------------------------

The PN.S astrometry can also be compared with SDSS astrometry.  Both
the differences in right ascension (RA) and in declination (Dec) for
our 66 PNe in common with M06 are extremely small ($\Delta {\rm
RA}$(Our$-$M06) = $0.004\pm0.9$ arcsec and $\Delta {\rm
Dec}$(Our$-$M06) = $0.08\pm0.8$ arcsec) and do not show any systematic
effects.

\begin{figure}[th]
\begin{center}
\includegraphics[width=10.0cm,angle=-90,clip=true]{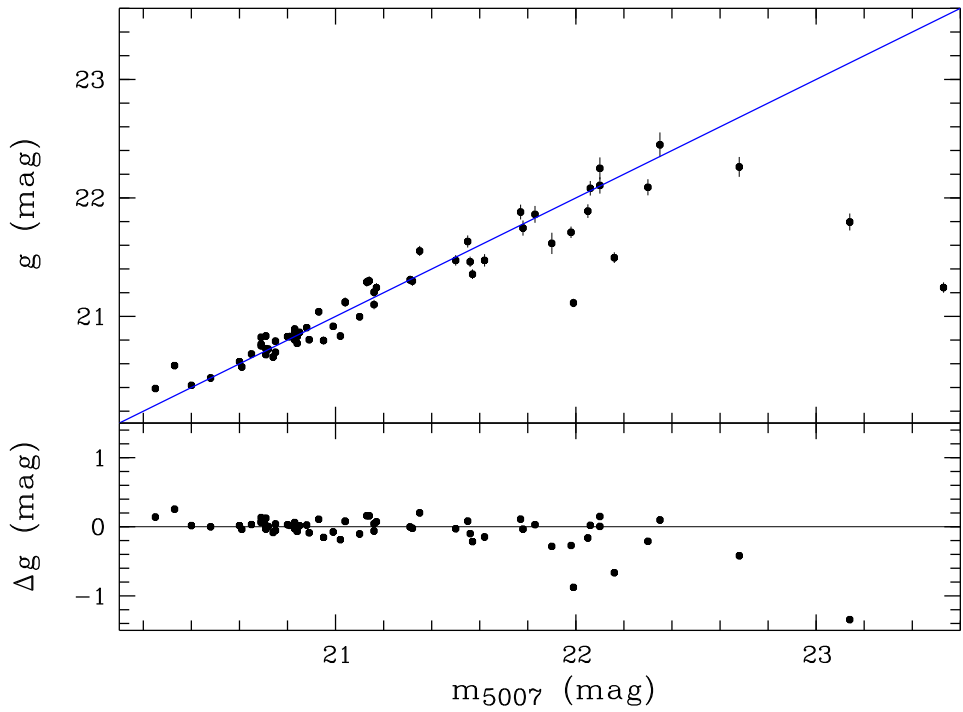}
\caption{\label{fig:g_v_5007}
Comparison of $g$ magnitudes and m$_{5007}$ magnitudes from M06.
{\it Top}: The solid line represents a line
of slope unity where the $g$ magnitudes equal the m$_{5007}$ magnitudes.
{\it Bottom}: Difference between $g$ and m$_{5007}$ magnitudes.
}
\end{center}
\end{figure}

Data from the SDSS together with data from M06 can be used to check
our basic hypothesis that the SDSS $g$ magnitudes for PNe are very
close to $m_{5007}$ magnitudes. Figure~\ref{fig:g_v_5007} shows a
comparison between $g$ magnitudes from the SDSS without extinction
correction and $m_{5007}$ magnitudes from M06 for 66 PNe that are
common to both samples.  As can be seen in the top panel of the
figure, most of the points are located around the line of slope unity
where both values would equal each other.  To check more accurately
for possible systematic differences, the value $\Delta
$($g-$m$_{5007}$) is drawn in the bottom panel.  As this panel shows,
the difference does not reveal any systematic trends up to
$g\sim22.5$~mag, which is around our detection limit.  

Only five data points show an underestimation of their flux in the
PN.S data starting from m$_{5007}\sim22.0$~mag, or an overestimation
of the $g$ magnitude by the SDSS.  Since the difference amounts to up to 2.3
mag, but all these PNe were observed with the 2.2m telescope, it seems
more likely that the flux was underestimated by M06.  Without these
five points the $g$ and $m_{5007}$ magnitudes agree within a
weighted standard deviation of 0.09 mag, which supports our assumption
of the approximate equivalence of the $g$ and $m_{5007}$
magnitudes.

%---------------------------------------------------------------
\section{The Efficiency of the Method}
\label{txt:efficiency}
%---------------------------------------------------------------

%***************************************************************
\subsection{Spectral observations}
%***************************************************************

\begin{figure}[th]
\begin{center}
\includegraphics[width=9.0cm,angle=-90,clip=true]{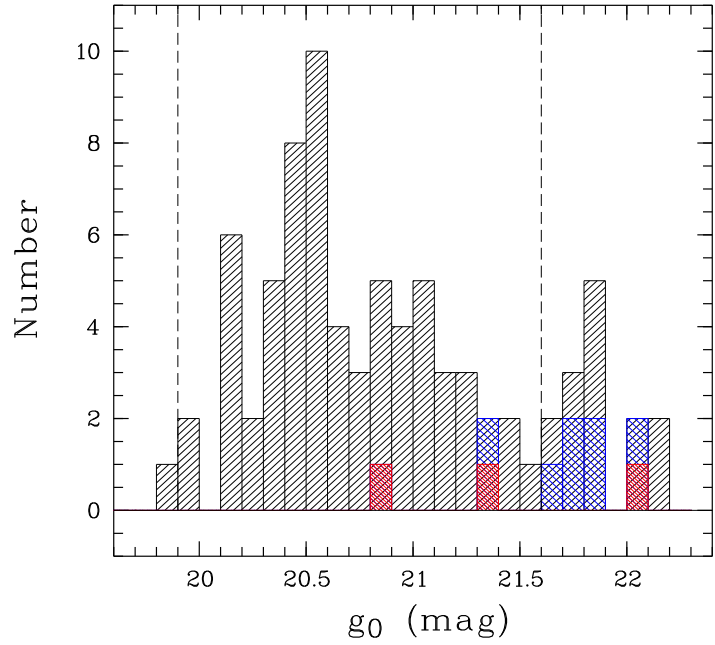}
\caption{\label{fig:Obs_hist}
Distribution of observed PN candidates in the $g_0$ band.  Blue double
hashed bins indicate the distribution of observed PN candidates
without obvious emission lines.  Red double hashed bins indicate
positions of three candidates that were additionally identified as
genuine PNe during cross-identification work.  The data are binned
into 0.1 mag intervals.  The magnitude criteria for the second priority
candidates selection are shown with vertical lines.
}
\end{center}
\end{figure}

Our spectroscopic follow-up observations of part of our sample allow
us to estimate the efficiency of our color-selection method.
Altogether, out of the 80 observed PN candidates in the SDSS M31 data,
70 objects turned out to be genuine PNe, resulting in an estimated
detection efficiency of $\sim88\%$.  The efficiency is different for
the first ($\sim95\%$) and the second priority ($\sim63\%$) candidates
and depends obviously on the magnitude and color of the selected
candidates.  The histogram distribution of the PN candidates as a
function of magnitude in the $g$ band is shown in
Figure~\ref{fig:Obs_hist}.  Cross-hatched bins indicate observed PN
candidates that turned out not to be PNe in our follow-up spectroscopy
(no obvious emission lines).  As can be seen from this figure the
detection efficiency is essentially 100\% for magnitudes brighter than
$g_0 = 21\fm6$, but shows a pronounced decrease for fainter
magnitudes.

There are two possibilities inherent to our method and data set that
help to explain this trend with luminosity.  Firstly, the decreasing
number of detected true PNe with decreasing luminosity may reflect the
increasing photometric errors for fainter magnitudes and the thus
increasing number of false detections.  This affects in particular the
SDSS $u$ band, since this band has the lowest sensitivity of all the
passbands used in our detection method.  At the same time, the
incompleteness of true detections is likely to increase towards
fainter magnitudes, since especially PNe with weak emission lines may
remain unrecognized in the photometric data.
Secondly, the small telescope employed for our follow-up observations
contributes to the difficulty of confirming fainter candidates.  There
is at least one PN candidate with $g = 22\fm0$, which we re-observed
with an exposure time of 1800~s after a 900~s exposure did not reveal
emission lines.  In the longer exposure emission lines were detected.
Furthermore, three additional, known PNe from the lists of HK04 and
M06 do not show any emission in our spectra (even though two of them
lie within the selection box of our ``first priority'' objects -- see
Figure~\ref{fig:Obs_hist}).  These three PNe are shown in
Table~\ref{tbl-1} with a flag value ``2'', and our total efficiency is
$\sim91\%$ ($\sim98\%$ for the first priority candidates and
$\sim68\%$ for the second) after taking them into account.

Hence we cannot be certain that our faint PN candidates without
detected emission lines in their spectra are indeed false
identifications -- deeper observations or observations with larger
telescopes may uncover weak emission lines after all and may thus
improve our detection statistics.  In this sense, our listed numbers
may, in fact, only be lower limits.  Therefore we did not remove
these candidates from Table~\ref{tbl-1}.

As explained in Section~\ref{txt:Sel_determ}, very red $(u-g)_0$
colors of PNe are caused by the strong emission line [O\,{\sc iii}]
$\lambda$5007 in the wavelength range covered by the $g$ filter and by
the absence of any strong emission lines in the $u$ band.  The
H$\beta$ emission line is also located in the $g$ filter, close to the
position of [O\,{\sc iii}] $\lambda$5007, but the ([O\,{\sc iii}]
$\lambda$5007/H$\beta$) line ratio is usually much stronger for
spectra of PNe as compared to spectra of ELGs.  This contributes to
the efficiency of our color-selection method to select PNe as opposed
to ELGs.

\begin{figure}[th]
\begin{center}
\includegraphics[width=10.0cm,angle=-90,clip=true]{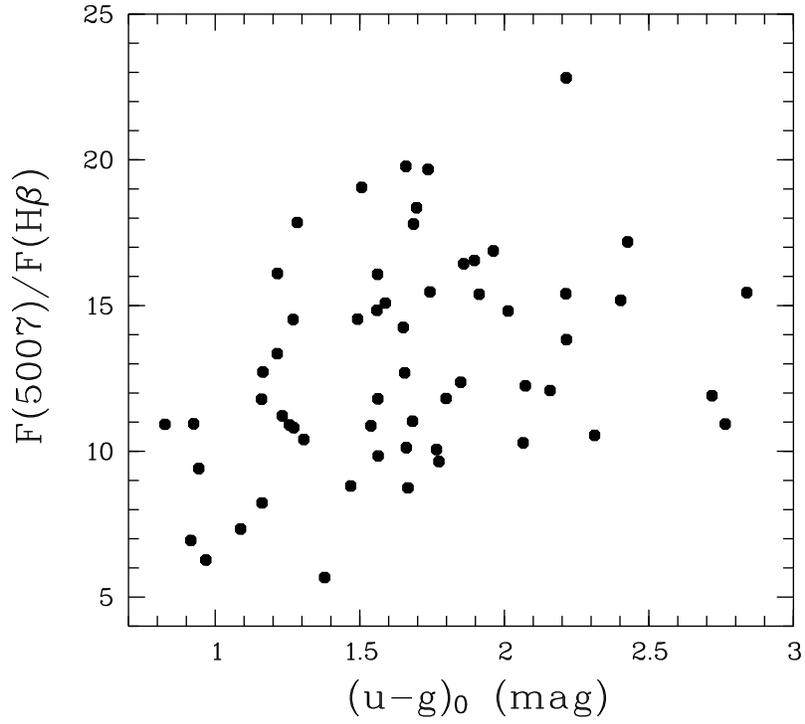}
\caption{\label{fig:5007_ug}
Distribution of observed [O\,{\sc iii}] $\lambda$5007/H$\beta$ line
ratio versus $(u - g)_0$ color to illustrate the working area of our
color-selection method.
}
\end{center}
\end{figure}

Using our observational data, we can try to evaluate our $(u-g)_0$
color criterion in terms of this ratio.  The distribution of the
observed ([O\,{\sc iii}] $\lambda$5007/H$\beta$) line ratio versus the
$(u - g)_0$ color is shown in Figure~\ref{fig:5007_ug}.  Taking into
account that ELGs with strong emission lines have a mean line ratio of
4.06$\pm$1.11 \citep{Kni04}, we can conclude that $(u - g)_0 =
0\fm6-1\fm0$ is approximately the limit where ELGs with strong lines
and PNe start to be comparable in SDSS colors.

%***************************************************************
\subsection{Cross-identification with other data}
%***************************************************************

\begin{figure}[th]
\begin{center}
\includegraphics[width=12cm,angle=-90,clip=true]{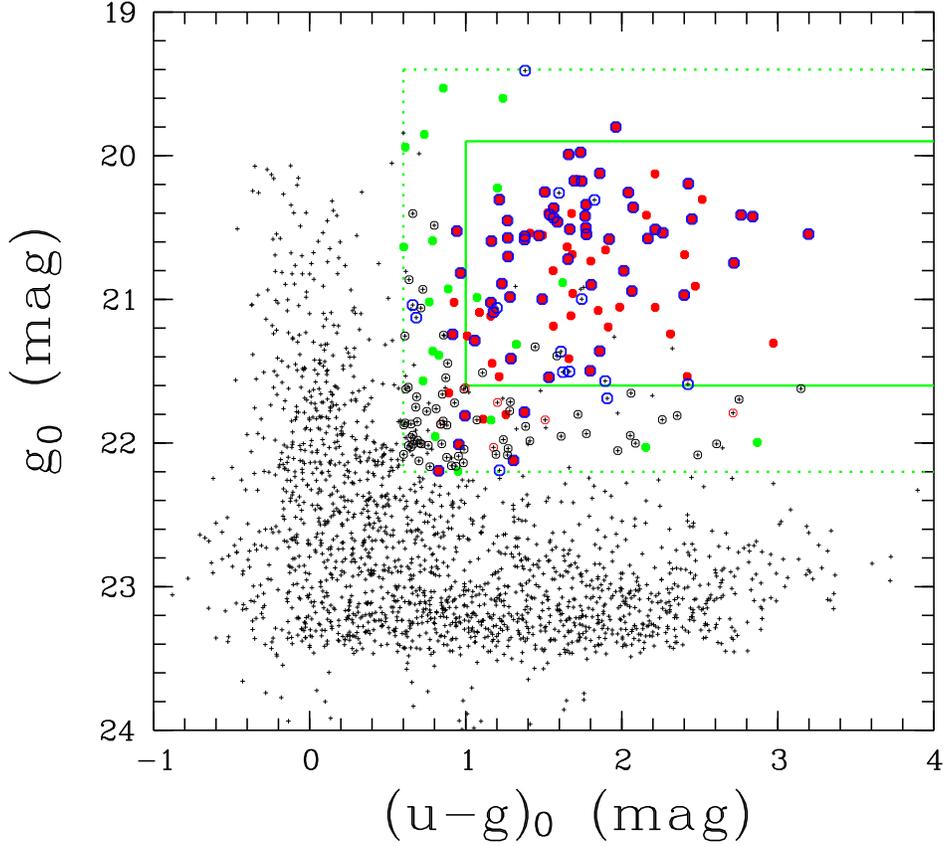}
\caption{\label{fig:obs1}
Same as the right bottom panel of Fig.~\ref{fig:Sel_crit}, but
results of observations, cross-identifications with samples from
HK04 and  M06 and visual checks using images of SLGG are also shown:
(1) all PN candidates are shown as crosses in the selected color-magnitude area;
(2) all PNe from the test sample and true PNe confirmed with spectroscopic
follow-up observations are shown as filled red circles;
(3) all observed PN candidates without
obvious emission lines are shown with empty red circles;
(4) all PNe that also were identified in the HK04 and/or M06 samples are shown
with empty blue circles;
(5) all non-observed PN candidates that were not identified in the 
HK04 and/or M06
samples, but that showed obvious emission in [O\,{\sc iii}] $\lambda$5007 and H$\alpha$ images
in the SLGG data are shown with green filled circles;
(6) all PN candidates that were deleted from the sample after a
cross-check with the SLGG data
are shown with empty black circles.
}
\end{center}
\end{figure}

Additional cross identifications with data from HK04 and M06 (73
identifications in total) and visual checks using images of the SLGG
(22 identifications in total) provide further possibilities to
evaluate our method.  
A CMD of the SDSS \MA\ data with all our PN
candidates and those currently known as genuine PNe is shown in
Figure~\ref{fig:obs1}.  All PN candidates are shown as crosses in
the selected color-magnitude area.  All PNe from the test sample and true
PNe confirmed with spectroscopic follow-up observations are shown with
red filled circles.  All PNe that were identified in the HK04 and/or
M06 samples are marked with empty blue circles.  All PNe candidates
that were not observed or not identified by HK04 and/or M06, but that
showed obvious emission in the [O\,{\sc iii}] $\lambda$5007 and
H$\alpha$ images in the SLGG data are shown with green filled circles.
Observed PN candidates without obvious emission lines that are still
in the sample are shown as empty red circles.  All PNe candidates that
were deleted from the sample after the visual inspection of the SLGG
data are depicted by empty black circles.

In total, 103 candidates were selected as first priority candidates,
out of which three were rejected after checking the SLGG data. Only
five first priority candidates still remain to be confirmed.
Altogether, this results in an efficiency of our method in the area of
the first priority candidates of at least 92\%.  The efficiency for
the area of the second priority candidates drops from $\sim100\%$ for
magnitudes $g<20\fm0$ to $\sim30\%$ for magnitudes $g>21\fm6$.

Finally, we made one more cross-identification search using the HK04
and M06 sample, but this time for {\em all} available \MA\ SDSS data
(both stellar and extended sources).  We identified only one
additional PN as compared to those previously identified in our
sample.  We thus conclude that with our selection criteria we are able
to select 99\% of the known PNe in the SDSS \MA\ data.

\begin{figure}[th]
\begin{center}
\includegraphics[width=10.0cm,angle=-90,clip=true]{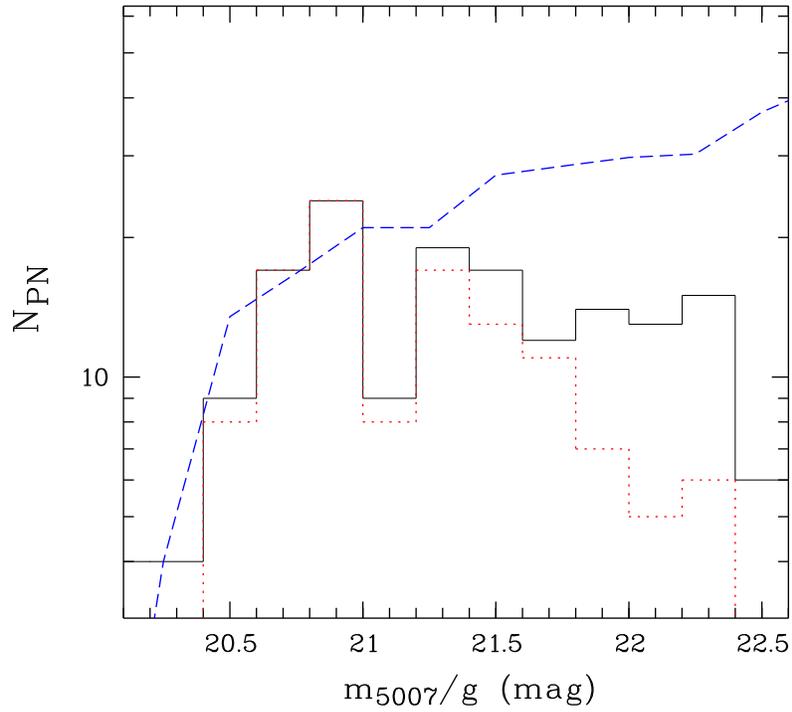}
\caption{\label{fig:PN_5007}
Comparison of the PN luminosity function (PNLF) from M06 with the PNLF
from this work.  The PNLF from M06 (m$_{5007}$ magnitudes) is shown
with the blue short dash line.  The PNLF based on all our selected candidates
($g$ magnitudes) is shown with the black solid line.  The PNLF for all
currently known genuine PNe from our sample is shown with the red dotted
line.  All data were binned into 0.25 mag intervals.
}
\end{center}
\end{figure}

We show a comparison of the PNLF from our work with the PNLF from M06
in Figure~\ref{fig:PN_5007}.  All data were binned in 0.25 mag
intervals.  The data were not corrected for reddening.
%PNLF for M06 is corrected for the extinction 0\fm32 in $g$ filter (it
%is median value for our sample) and normalized (blue line).
The PNLF from M06 is plotted as a blue line.  The black line shows the
PNLF for our sample assuming that all selected PN candidates are real
PNe.  The red line shows the PNLF for confirmed PNe from our sample.
As expected for a presumably universal curve regardless of the region
sampled, at the bright end our PNLF and the one of M06 show excellent
agreement.  Our completeness limit is about $g = 21\fm0-21\fm2$.

%---------------------------------------------------------------
\section{A Few Characteristics of the Planetary Nebulae in M31}
\label{txt:M31PNe}
%---------------------------------------------------------------

%---------------------------------------------------------------
\subsection{Spatial Distribution}
\label{txt:space}
%---------------------------------------------------------------

\begin{figure*}[th]
\begin{center}
\includegraphics[angle=0,width=18.0cm,clip=true]{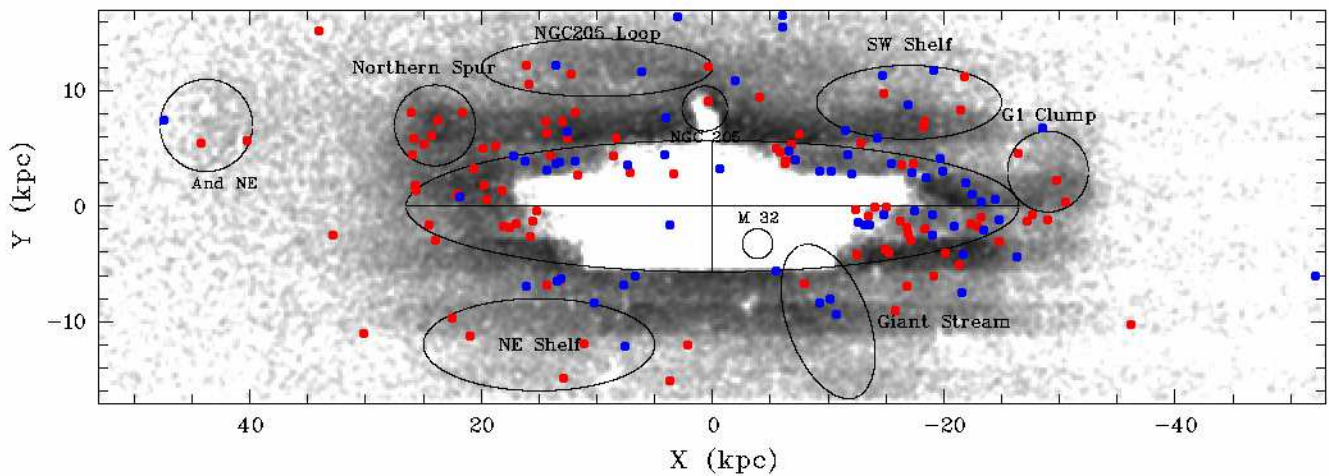}
\caption{\label{fig:Obs_final}
Spatial distribution of newly-discovered PNe from the \MA\ SDSS data.
The spatial distribution of all stars that were detected by the SDSS
standard pipeline in M31 data is shown. The data were binned by
$2\arcmin \times 2\arcmin$.  $X$ and $Y$ are in kpc from the center of
\MA\ along the major and minor axes, respectively.  The central
ellipse has a semimajor axis of 2 degrees ($\approx27$~kpc) and
represents the optical disk of \MA\ with $i=77.5$, which lies well
within this boundary.  First priority candidates are shown with red
filled circles and second priority candidates are shown with blue
filled circles.  The other ellipses mark the locations of NGC\,205,
M32, Andromeda NE, and some other major substructures shown by
\citet{Fer02,Fer05,Ibata07,Rich08}.
}
\end{center}
\end{figure*}

The spatial distribution of all newly discovered PNe is shown in
Figure~\ref{fig:Obs_final}. They are overplotted on top of all stars
detected by the SDSS. It is obvious that the discovered PNe really
trace the distribution of stars in the outer regions of \MA.  In part,
many of them are seen superimposed on various of the recently
uncovered well-known morphological features like the Northern Spur,
the NE Shelf, the NGC\,205 Loop, the G1 Clump, etc. For a more
detailed description of these features and of their stellar
populations, see \citet{Fer02,Fer05,Ibata05,Ibata07,Rich08}.  

Certain structures stand out in the spatial distribution of the PNe: A
large number of PNe is seen in the area of the Northern Spur and three
PN candidates are identified at the location of Andromeda NE
\citep{Zucker04a}.  Those newly discovered PNe, which could be
associated with known structures in the outer regions of \MA, are
marked in Table~\ref{tbl-1}.  However, whether individual PNe are
associated with M\,31's disk itself or alternatively with tidal
streams (or other \MA\ components) cannot be decided solely based on
their location or spatial coincidence.  Such examples will be
discussed in the next sections.

There is a certain asymmetry in the distribution of newly detected PNe
around \MA: we find more in the upper and central portion of
Figure~\ref{fig:Obs_final} (corresponding to the northwestern part of
\MA) than in the lower portion.  Hardly any PNe appear to be
associated with the giant stellar stream, while there are several in
the area of the NE shelf.  \citet{Fer05} point out that the stellar
populations of the NE shelf and of the Giant Stream are very similar,
but that the Giant Stream is approximately 60 kpc more distant from us
than the shelf.  \citet{Tan10} found the distance to the Giant Stream
to be even twice as large: 883$\pm$45 kpc in total.  Deeper data might
reveal more PNe in these two metal-rich features whose stellar
population properties appear to be so similar.  Alternatively, the
lower number of detected PNe could also be an effect of the PNLF
affected by the inclination of M31's disk relative to the observer
\citep{Mer06}, and a higher extinction in the more distant part as
seen from our perspective.

%---------------------------------------------------------------
\subsection{A minor axis density profile for \MA}
\label{txt:minor_axis}
%---------------------------------------------------------------

The very extended, diffuse, and disturbed outer regions of \MA, which
are visible in, e.g., SDSS star counts as shown in
Figure~\ref{fig:Obs_final}, is not a unique occurence.  For instance, deep optical
surveys of nearby face-on and edge-on dIrrs and disk galaxies have also
found that their stellar distributions are much more extended than
previously thought, e.g., NGC\,6822 \citep{deBlok06}, Leo\,A
\citep{Vans04}, IC\,1613 \citep{Batt07}, Pegasus \citep{Kniaz09}, the
Magellanic Clouds \citep[][and references therein]{Casetti-Dinescu12}
and the works of \citet{Malin99,Ti05,Ti06}.  Moreover, warps and
flaring have been shown to be common features of galactic disks
\citep[e.g.,][and references therein]{Guijarro10,vanderKruit11}. In
addition,  accretion features are now commonly detected
\citep[e.g.,][]{Delgado08,Delgado09,Delgado10,Mouhcine10,Miskolczi11,Ludwig12}.
In the case of \MA\ we have the opportunity to use PNe as tracers of
intermediate-age stellar components in these various structures.

In the case of \MA\ we appear to see an extension of the disk, which
has been shown to possess a complex structure with considerable
warping, both in the optical and in H\,{\sc i}
\citep[e.g.,][]{Brinks84,Walterbos88,Braun91,Morris94,Carb10},
probably caused and modified by interactions
\citep[e.g.,][]{McConnachie09,Richardson11,Qu11}.
In recent years the surroundings of \MA\
have been mapped using deep ground-based photometric
surveys of the Andromeda galaxy
\citep{Fer02,Irwin05,Ibata07,McConnachie09} with the wide-field
cameras of the Isaac Newton Telescope (INT) and the Canada France
Hawaii Telescope (CFHT).  \citet{Ibata05,Ibata07} presented a surface
brightness profile for \MA\ and concluded that along the minor axis,
in the region $0.2^\circ<R<0.4^\circ$, the classical inner (thin) disk
of \MA\ contributes to the profile, but at $0.5^\circ<R<1.3^\circ$ the
extended disk component becomes dominant.

Since PNe trace the distribution of the underlying intermediate-age
stellar populations, we construct a PN density profile for \MA\ using
the catalogue of PNe we compiled from our sample and from the samples
of HK04, H06, and M06 (see Section~\ref{txt:compar}).  We used the
same method as in \citet{Kniaz09}, where the density profile for the
outer parts of the Pegasus dIrr was constructed using star counts: PN
densities were calculated within elliptical apertures with a fixed
aspect ratio.  The following assumptions and rules were used during
this procedure: (1) We assume that the PNLF of \MA\ does not vary
throughout the entire \MA\ area. (2) For the central part of \MA\ ,
i.e., the region inside of the classical disk (within an ellipse with
a semimajor axis of two degrees in Figure~\ref{fig:Sel_pos1}), where
the standard SDSS software does not work properly due to crowding, the
sample from M06 provides most of the PNe and our data added only very
few objects.  The aspect ratio for this region was chosen to be the
same as for the optical disk.  All PNe from the compact elliptical
galaxy M\,32 were excluded.  The PN densities were calculated in
elliptical apertures with a stepwise axis increase of 0.01 degrees.
(3) For all PNe outside of this inner region we first limited the
sample to PNe brighter than $g = m_{5007} = 22.5$~mag (see
Figure~\ref{fig:PN_5007}). In addition, all PNe from the dwarf elliptical
galaxy NGC\,205 were excluded.  An aspect ratio of 3:5 was used for
this region \citep{Fer02,Irwin05,Ibata07}.  PN densities were
calculated within elliptical apertures with a step of 0.2 degrees.
The PN data cover only part of the studied region and this geometrical
incompleteness increases with radius. To correct for this effect, we
used an ``incompleteness factor'' -- the ratio between the total area
for the elliptical annulus and the actually covered area.  This factor
is about 1.0 at $0.5^\circ<R<1.0^\circ$, but increases drastically
after that.  We included this factor in the error propagation. All
subsequent fitting was done with weights of $w_k = \sigma_k^{-1}$,
where $\sigma_k$ is the uncertainty calculated for each level
\citep{LSBs}.

\begin{figure}[th]
\begin{center}
\includegraphics[width=12.0cm,angle=-90,clip=true]{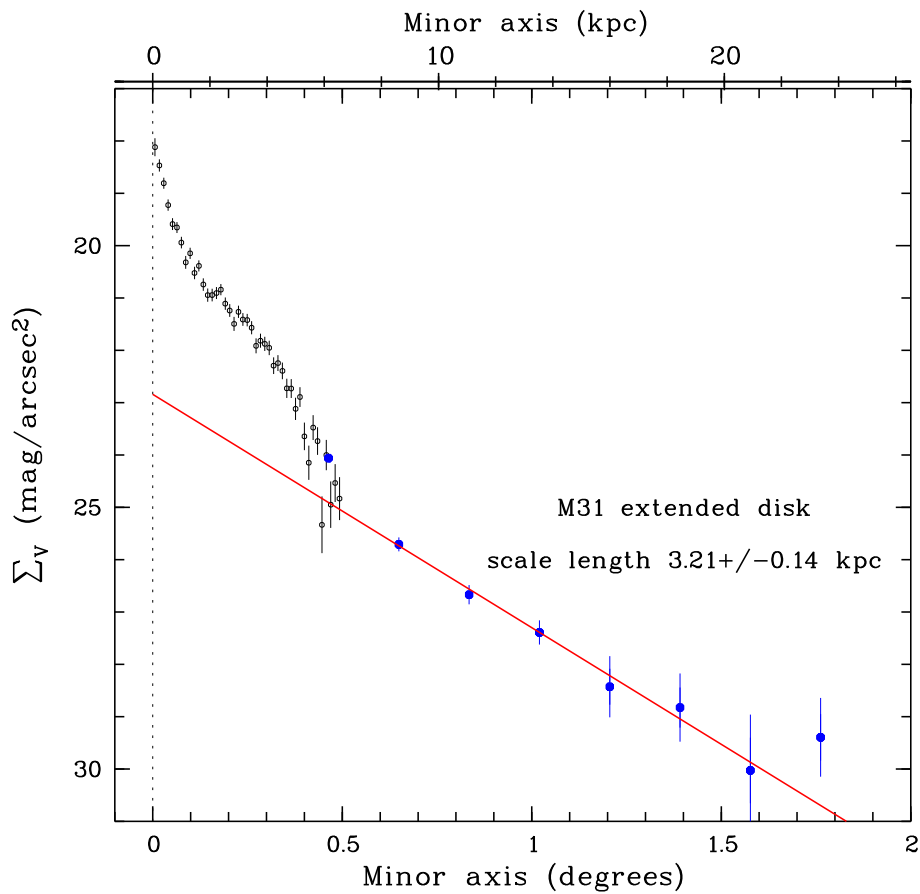}
\caption{\label{fig:exp_disk}
Calculated distribution of the PN density along the minor axis.  Data
for the central part of \MA\ are shown with small empty circles, while
data for the outer part are shown with blue circles.  All PN densities
were calculated within elliptical apertures in steps of 0.01 degrees
for the central part and 0.2 degrees for the outer part.  The final
density distribution was normalized to have $\sim25$ mag arcsec$^{-2}$
at a distance of 0.5$^\circ$ from the center.
Fitting the data in the region $8<R<20$~kpc yields a scale length of
3.21$\pm$0.14~kpc that is similar to the scale length
3.22$\pm$0.02~kpc for the same region calculated by \citet{Ibata07}
using photometric data from the Isaac Newton Telescope Wide Field
Camera survey of \MA.
}
\end{center}
\end{figure}

The final density distribution was normalized such that the surface
brightness level at a distance of 0.5$^\circ$ from the center of \MA\
is $\sim25$ mag arcsec$^{-2}$.  This normalization was chosen in order
to be comparable with the V-band minor-axis surface brightness profile
shown in Figure~51 of \citet{Ibata07}.  Our calculated density
distribution along the minor axis is shown in
Figure~\ref{fig:exp_disk}.  Data calculated for the central part of
\MA\ are marked by small open circles, and data for the outer part are
shown as blue circles.  Error bars indicating the uncertainties for
each point are also plotted.

Our profile looks very similar to the $V$-band minor-axis surface
brightness profile of \citet[][their Figure 51]{Ibata07}: in the very
center at minor axis radii $R<0.1^\circ$ we clearly see the bulge.
Farther out in the region of $0.1^\circ<R<0.5^\circ$ the classical
inner disk (with bumps caused by spiral arms) contributes to the
profile.  Then the extended disk becomes dominant, and our profile
shows an exponential decline out to the $R=20$~kpc, where our data, in
principle, still trace the extended disk.  Fitting the data for the
region $8<R<20$~kpc, we measure an exponential scale length of
3.21$\pm$0.14~kpc, which is quite similar to what was found for the
same region by \citet{Irwin05} and \citet{Ibata07}, who found a scale
length of 3.22$\pm$0.02~kpc along the minor axis profile using
photometric data from the INT Wide Field Camera survey of \MA.  This
structure has a very low central surface brightness at a level of
$\mu_0\sim23$~mag arcsec$^{-2}$.

Various scenarios for the formation of the extended disk were
discussed by \citet{Ibata05}.  These authors concluded that the most
probable scenario is the formation via accretion of many small
subgalactic structures.  \citet*{Pen06} interpreted the extended disk
as a possible result of a single dwarf satellite merger (with a mass of
$10^9-10^{10} M_{\sun}$) and suggested that the inner disk would not
be strongly affected by such an accretion event \citep[see
also][]{Hammer10}.  \citet{Magr05b} correlate the number of PNe within
four magnitudes of the absolute magnitude of the PNLF bright-end
cut-off with the total stellar mass for different galaxies (their
Fig.\ 5).  Taking the number of PNe detected in the outer region of \MA\
into account ($\sim$200), the fact that most of them belong to the extended disk,
and considering the relation derived by \citet{Magr05b}, we can
conclude that the PNe in this area trace a total stellar mass
comparable to that of M\,33 ($\sim10^{10} M_{\sun}$) or to the
stellar mass estimate inferred for M31's thick disk component as
analyzed by \citet{Collins11}. If we then assume that the
abundance distribution for PNe in the extended disk was also similar
to the distribution for M\,33, we would expect a mean oxygen abundance
value close to 12+log(O/H)=8.2--8.4 dex following \citet{Magr09b}.

Finally, we note that a similar, very extended (up to 40--50 kpc) and
rotationally supported disk-like structure was found with PN data
in the halo of the nearby peculiar giant elliptical galaxy
Centaurus\,A \citep[][]{PFF04}.

%---------------------------------------------------------------
\subsection{Extended disk or rotating spheroid?}
\label{txt:M31velocities}
%---------------------------------------------------------------

A number of studies of the \MA\ surface brightness profile and of the
stellar populations in \MA\ have suggested that fields as far out as 20~kpc
were still dominated by the bulge \citep[e.g.][]{PB94, Durrell01,
Irwin05}.  Deep HST imaging studies of selected minor-axis fields out
to 35 kpc suggest that both the spheroid and the giant stellar stream
contain, in part, similar (though not identical) intermediate-age
populations younger than 10 Gyr and that the spheroid populations are
polluted with stars from the progenitor of the stream
\citep[e.g.,][]{Brown06,Brown07,Brown08}. These photometric findings
are supported by the spectroscopic studies of red giants along the
minor axis by \citet{Gilbert07} and \citet{Fardal12}.  

Altogether, out to 20 kpc from the center of \MA\ we see the
superimposed contributions of stars belonging to \MA's spheroid (bulge
and halo components), disk components, and accreted components.   In
their kinematic study, \citet{Collins11} identify a dynamically hotter
thick disk component in \MA\ (in addition to the colder classical thin
disk and extended disk).  \citet{Dorman12}, in another kinematic study
based on red giants, find that although the disk components dominate
in the inner 20~kpc of \MA, the inner spheroid can be traced
throughout this region as a rotating, hot component.  

HK04 had a sample of 135 PNe that only cover a fraction of \MA, but
their data include a similar range of minor and major axis distances
from the center of \MA\ as the above studies that are based on red
giants.  HK04 carried out dynamical modeling and conclude that a
standard model for the thin and thick disks and bulge can not
reproduce the observed PNe kinematics.  They suggest that the majority
of the PNe in their sample are probably members of a very extended
bulge, which rotates rapidly at large distances and dominates over the
halo out to at least 20 kpc.  (In the more recent literature this hot
rotating component is commonly referred to as ``inner spheroid''
\citep[see, e.g.,][]{Dorman12,Fardal12}.

\citet{Ibata05} discovered an additional extended disk population and
reassessed HK04's data in the light of this discovery.  They concluded
HK04's data may favor the disk interpretation, but could not confirm
or exclude whether a rotating spheroid is still needed.  \citet{Mer06}
obtained velocities for a vast number of PNe and found that \MA's
extended bulge (or spheroid) component can be traced out to ten
effective bulge radii or approximately 15~kpc. \citet{Far07} conclude
that \citet{Mer06}'s global PNe kinematics can be best represented as
a combination of \MA\ components plus debris from accreted satellites,
including the counterrotating shelf structures.  

In their kinematic PN study, \citet{Hall06} argue that the
substantial drop in the PN velocity dispersion from the center of
\MA\ ($\sim 130$~km~s$^{-1}$) out to about 11~kpc ($\sim
50$~km~s$^{-1}$) along the major axis does not support the existence
of a dynamically hot PN halo, but they, too, find evidence for an
extended bulge component akin to HK04.  For the PNe belonging to the
disk component \citet{Hall06} find a rotation velocity of $\sim
140$~km~s$^{-1}$. As pointed out by \citet{Kwitter12}, the PN
velocity dispersions at large radii and those of the thick disk red
giants measured by \citet{Collins11} are roughly in agreement.

\begin{figure}[th]
\begin{center}
\includegraphics[angle=-90,width=18.0cm,clip=true]{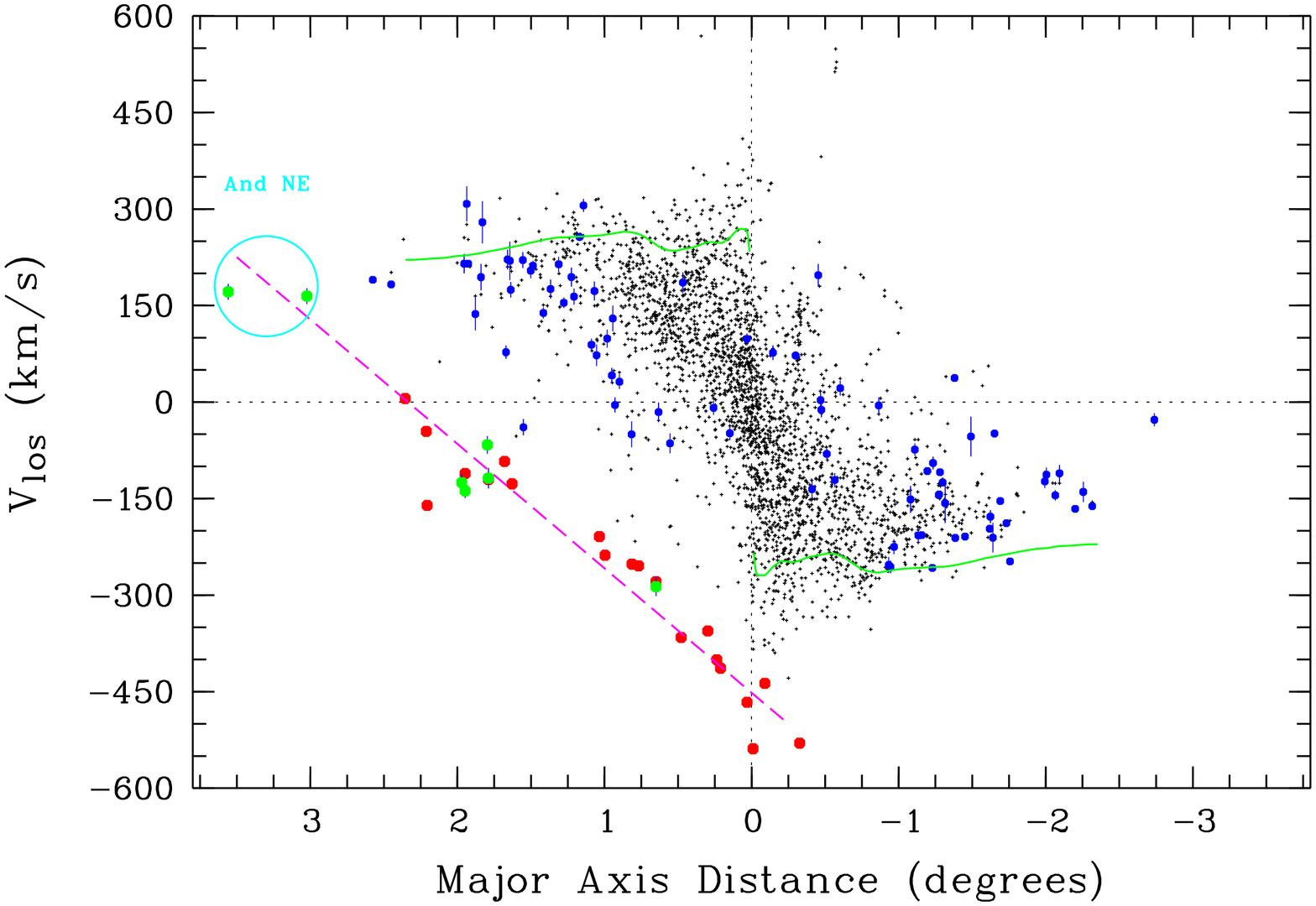}
\caption{\label{fig:gen_vels}
Velocity-distance diagram for newly discovered PNe in \MA.  All shown
velocities of PNe are corrected for the systemic velocity $V_{sys} =
-306$ km~s$^{-1}$ \citep{Carb10}.  $X$ represents the projected
distance along the major axis.  The line is the H\,{\sc i} rotation
curve from \citet{Kent89}.  Black dots indicate PN data from M06 and
HK04.  Filled blue and green circles are newly observed PNe from our
sample.  The uncertainties of the velocity determination are shown
by vertical error bars.  Red filled circles mark PNe from M06, which were
identified by M06 as forming a continuation of the Giant Stream.
Green filled circles show those of our new PNe suggested to belong to
the ``continuation of the Giant Stream''.  The position of PNe
associated with Andromeda\,NE is also shown.
}
\end{center}
\end{figure}

A full dynamical model fit to all PN data is beyond the scope of this
paper. But considering the data shown in Figures~\ref{fig:Sel_pos1},
\ref{fig:Obs_final}, \ref{fig:exp_disk}, and \ref{fig:gen_vels} we
suggest that the domination by an immense rotating spheroid traced by
PNe out to $R\sim20$~kpc along the minor axis looks less probable than
the extended disk scenario: (1) The spatial PN distribution follows
that of the stars with an aspect ratio close to 3:5 for the outer part
of \MA. (2) The density profile along the minor axis for the PNe is
very similar to the surface brightness profile of \MA\ for the region
$0^\circ<R<0.5^\circ$ and shows a similar exponential-like profile in
the region $0.5^\circ<R<1.5^\circ$. (3) As also found in the studies
mentioned earlier, clearly the bulk of the PNe does not show the
signature of a kinematically hot halo where all objects would have
random velocities except for the very central part dominated by the
bulge population; (4) most PNe (except for those in the very central
part) exhibit a distribution indicating that they belong to a
component that is rotationally supported.

In Figure~\ref{fig:gen_vels} we show the velocity distribution of all
PNe from HK04, H06, M06, and our sample along the major axis of \MA.
All shown velocities were corrected for the systemic velocity of \MA\
of V$_{sys} = -306$ km s$^{-1}$ \citep{Carb10}.  This plot shows that
the majority of the PNe belongs to the rotationally supported system,
where most of our new PNe (blue circles) have velocities
systematically shifted to lower values.  This may be the expected
situation in the case of, for example, a simple model for the velocity
of stars on circular orbits around \MA\ \citep{Ibata05}.
Alternatively, we may be seeing the difference between the kinematics
of the inner part of \MA\ and the extended disk as predicted by
\citet{Pen06}.  Which of these scenarios is the more likely one cannot
yet be answered with the existing data. 

There are also indications for the presence of an intermediate-age
population in the halo of \MA\ that may amount to up to 30\% of the
stars  \citep{Brown03}.  The few PNe that deviate from the rotational
signature might be members of the halo of \MA\ or of its ``extended
spheroid'' \citep[see discussion of the vast extent of \MA\
in][]{W05,Tan10}.  Considering the apparent presence of an
intermediate-age population in \MA's halo, PNe should exist there as
well, but since the number of PNe strongly depends on the luminosity
of the underlying population \citep[e.g.,][]{Ciardullo89} only a
handful of PNe can be expected.  However, whether the intermediate-age
population is indeed part of the halo is still being debated,
especially when considering that \MA's disk may be even more extended
than commonly thought, that the existing pencil-beam pointings may be
contaminated by stream stars, and that there are apparent large-scale
variations in the stellar populations in \MA's outer regions
\citep{Durrell04,Fer05,Chapman05}.  We find very few PNe with
distances of more than 30 kpc from \MA's center.

%---------------------------------------------------------------
\subsection{PNe in Andromeda NE}
\label{txt:AndNE_ab}
%---------------------------------------------------------------

We identified three PN candidates at the location of Andromeda NE
\citep{Zucker04a}.  The nature of this diffuse low-surface brightness
structure in the outer regions of \MA\ is still unclear.  It may be a
very low-mass and low surface-brightness galaxy, a portion of an
extended stellar tidal stream, or possibly just turn-off material from
the disk of \MA.  Andromeda NE has an absolute luminosity in the
$g$-band of $\sim -11.6$~mag and a central surface brightness of only
$\sim 29$ mag arcsec$^{-2}$ \citep{Zucker04a}.

According to \citet{RB86} the number of stars n$_P$ in any
post-main-sequence phase $P$ can be calculated with the equation
%\begin{equation}
$   n_P  = \eta\,L_T\,t_P$,
%\end{equation}
where $\eta$ is the number of stars per unit luminosity that leave the
main sequence per year (between 5$\times10^{-12}$ to 2$\times10^{-11}$
yr$^{-1}$ $L_{\odot}^{-1}$), $L_T$ is the total luminosity of the
galaxy (5$\times10^6 L_{\odot}$ for Andromeda NE), and $t_P$ is the
duration of the evolutionary phase ($\le$~20,000 years for PNe).
Applying this formula shows that the number of PNe that might be
expected in Andromeda NE is 0.5--2.  These numbers are in good
agreement with the number of PN candidates that we found in this area.
Both our 2.2m spectroscopy in this paper and later 3.5m (Calar Alto,
Spain) and 6m Russian telescope spectroscopy (Kniazev et al., in
preparation) confirm that all these candidates are real PNe, which
could be used for kinematical and chemical studies of Andromeda NE.
Two of the PNe presented in this work are located at projected
distances of $\sim$48~kpc and $\sim$41~kpc from the center of \MA\ and
are the most distant PNe in \MA\ found up to now.

\citet{Ibata05} obtained stellar spectra of presumed \MA\ disk stars
with the Keck DEIMOS multi-object spectrograph.  Only one of their
DEIMOS fields ($16.7\times5$ arcmin) was located in the area of
Andromeda NE. The measured heliocentric velocities for 92 stars in
this field show a narrow distribution in the region $V_h=-100$ to
$-200$~km~s$^{-1}$.  This velocity distribution peaks at about
$V_h=-150$~km~s$^{-1}$ \citep[Figure~24 in][]{Ibata05} though there
may be some contamination from Galactic foreground stars.  All our
newly found PNe in the area of Andromeda NE have heliocentric
velocities in a very narrow velocity range close to
$V_h=-150$~km~s$^{-1}$, which suggests that Andromeda NE has an
average velocity close to that value.

%---------------------------------------------------------------
\subsection{Could Andromeda NE be the core or a remnant of the Giant Stream?}
\label{txt:AndNE_core}
%---------------------------------------------------------------

In Figure~\ref{fig:gen_vels} we show all PNe identified as a possible
continuation of the Giant Stream by \cite{Mer03}.  The Giant Stream
was first detected by \citet{Ibata01} at the southeastern outer part
of \MA, close to the minor axis, and was photometrically and
spectroscopically studied at various locations along the stream by,
e.g., \citet{McC03,Ibata04,Guha06}.  The continuation of this stream
in the internal part of \MA\ is rather uncertain.  It is also unknown
whether the progenitor of the stream has survived, and if so where it
is. For these reasons various possible scenarios have been suggested and
studied and different models for the Giant Stream have been
calculated \citep[e.g.,][]{Mer03,Fer02,Ibata04,Far06,Far07,Far08}.

To constrain the stream's orbit \citet{Far06} carried out N-body
simulations.  They find that the PN distribution of \citet{Mer03} can
be fit well when assuming that it is part of the expected extension of
the stream. They also note that it is easy to make the orbit of this
stream pass through Andromeda NE but warn that there is no strong
evidence that Andromeda NE is indeed part of the stream since it is
not visibly connected with it.  

In Figure~\ref{fig:gen_vels}, we plot all our PNe as green circles
that are located within the same region of ($X$, $V_{\rm los}$)
parameters as the previously identified PNe that may represent a
continuation of the Giant Stream \citep[see][]{Far06}.  Three of our
PNe were identified before, and our velocities for them are very close
to the values of \citet{Mer06}.  Two of our PNe are new. It is easy to
see that positions of the two new PNe in Andromeda NE are located
along the extension of the line that goes through the ``continuation
of the Giant Stream'' PN sample.

\begin{figure}[th]
\begin{center}
\includegraphics[angle=-90,width=15.0cm,clip=true]{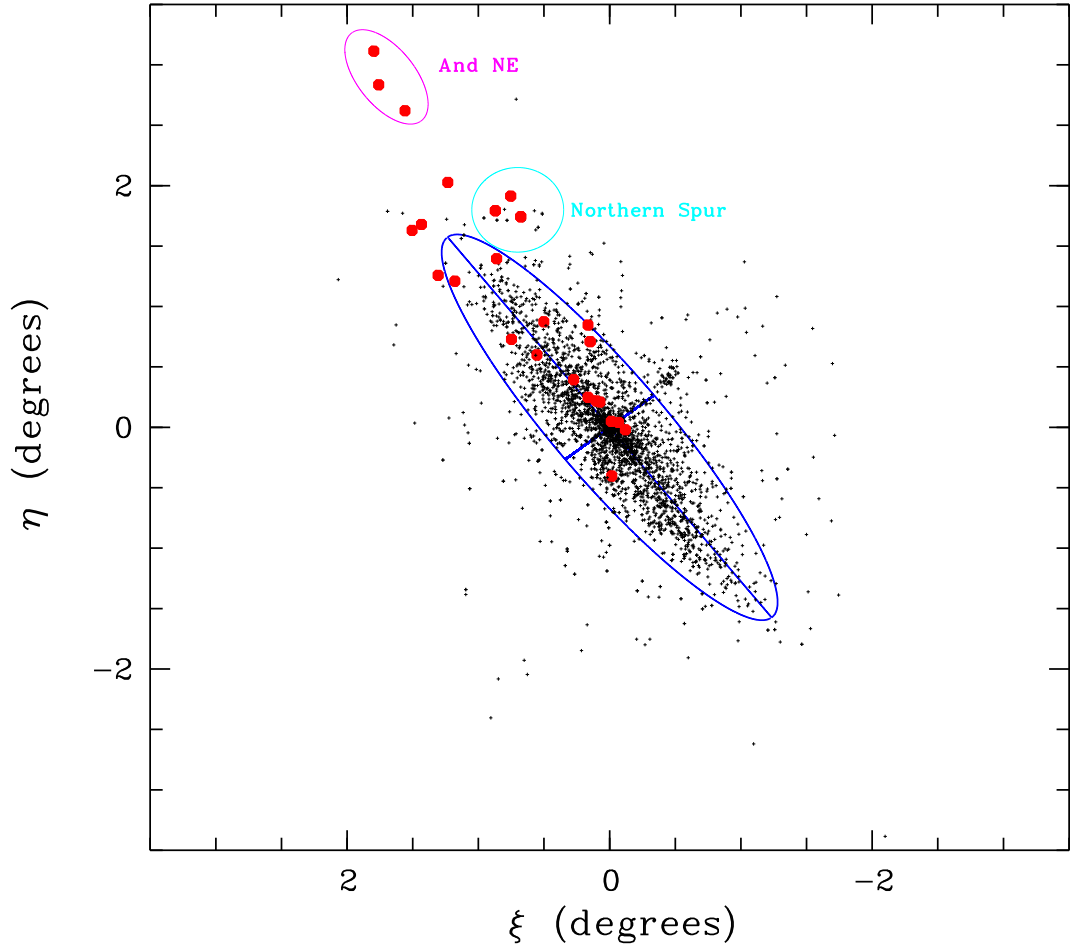}
\caption{\label{fig:gen_stream}
Positions of all PNe from HK04, M06, and our sample relative to the
center and orientation of \MA\ (see caption of
Figure~\ref{fig:Sel_pos} for more details). Red filled circles show
PNe from M06 and our sample, which were identified as forming a
continuation of the Giant Stream.  The positions of the Northern
Spur and Andromeda\,NE are shown.
}
\end{center}
\end{figure}

In Figure~\ref{fig:gen_stream} we also plot positions of all these PNe
relative to the center and orientation of \MA.  All PNe from the
``continuation of the Giant Stream'' sample show a cone-like
structure, beginning close to the southeastern minor axis area, then
expanding to northeastern direction and covering the Northern Spur and
Andromeda NE area.

Taking into account the results of \citet{Far06} and our own plots, we
suggest that the PN data support the notion that Andromeda NE could be
a remnant of the Giant Stream.  If this suggestion is correct, then
there are altogether 24 PNe in this putative stream sample.  

Using PN surveys of (other) LG galaxies, \citet{Magr05b} infer a
relation between the number of PNe and the stellar mass of the
intermediate-age population of a galaxy.  Applying their relation, we
estimate that the Giant Stream progenitor had a mass of $\approx10^9
M_{\sun}$ \citep{Magr05b}, close to the mass of the NGC\,205 dwarf
elliptical galaxy.  Similarly, using the relation between the number
of PNe in a given galaxy and its $V$-band luminosity derived by
\citet{Magr03}, we estimate a total luminosity $10^8-10^{8.5}
L_{\odot}$ for the progenitor of the Giant Stream.  Comparing this
value to its currently measured luminosity, we suggest that about 90\%
of its stars have been lost during the interaction with \MA.  This
estimate is very close to the calculated dynamical mass of the Giant
Stream progenitor, $M_s \approx 10^9 M_{\sun}$, from \citet{Far06}.

Our estimate of the possible luminosity range of the progenitor is of
the order of the luminosities of the dwarf elliptical companions of
\MA\ as well as the Local Group dwarf irregular galaxy IC\,10.  The
estimated lower limits of the metallicity of the PNe in NGC\,185 and
NGC\,205 are $12+\log$\,(O/H) = 8.2 and 8.6, respectively
\citep{Richer95}.  For the PNe in NGC\,147 \citet{Goncalves07} find a
mean metallicity of $12+\log$\,(O/H) $=8.06^{+0.09}_{-0.12}$.
In IC\,10, only one PN metallicity based on a direct
measurement of $T_{\rm eff}$ has been published so far, yielding
$12+\log$\,(O/H) = 7.96 \citep{Magr09a}.  The lower limits estimated
for the remaining PNe without direct $T_{\rm eff}$ measurements are
higher.  We recall that for a given luminosity early-type dwarfs tend
to have higher metallicities than late-type dwarfs by up to 0.5 dex
\citep[e.g.,][]{Richer99, Grebel03}.  In any case, regardless of the 
morphological type of the progenitor, these sparse data
on known or estimated PN abundances in dwarf galaxies in the infered
luminosity range of the Giant Stream progenitor suggest that the
intermediate-age stellar populations in the progenitor may have had  
metallicities $12+\log$\,(O/H) of the order of 8.0 to 8.6 dex.

Finally, we would like to emphasize two findings resulting from our PN
analysis:  (1) The estimated difference in mass between the Giant Stream
progenitor ($\approx20$ PNe; $\sim10^9$~M$_{\sun}$) and the extended
disk of \MA\ ($\approx200$ PNe; $\sim10^{10}$~M$_{\sun}$) is an
order of magnitude.  For that reason the extended disk cannot be the
result of a merger of the Giant Stream progenitor and \MA\ as a number
of models have suggested \citep[e.g.,][]{Far06,Far08}. (2) Some part
of the Giant Stream progenitor may be spread across the extended disk,
but its contribution has to be small compared to the material of the
extended disk itself.

%---------------------------------------------------------------
\subsection{The Giant Stream and the Northern Spur connection}
\label{txt:GS_NE_connect}
%---------------------------------------------------------------

The Northern Spur was first noticed by \citet{Walterbos88}.  This
structure is located near the northeastern major axis of \MA.  The
direction of the gaseous warp \citep[e.g.,][]{Carb10} provides strong
support for the association of the Northern Spur with a warp in the
outer stellar disk.  \citet{Fer02} were the first to show that this
feature is an excess of stars at the same distance as \MA\ and that it
is about a factor of 1.5--2 times more overdense than the G1 clump.
They also suggested that the Northern Spur possibly consists of
intermediate-age stars of moderate metallicity, [Fe/H]$\ge-0.7$ dex.
Analyzing a DEIMOS field ($16.7\times5$ arcmin) in this area
\citet{Ibata05} found that the stellar velocities are similar to the
velocities of other fields located in the outer part of \MA.
\citet{Fer02, Zucker04a} and \citet{Ibata07} suggested that the
metallicity of the Northern Spur agrees well with those of  many other
parts of the extended disk of \MA\ such as the G1 clump and Andromeda
NE.  \citet{Rich08} analyzed HST data of a number of fields sampling
different areas of the outer part of \MA\ and distinguished two types
of fields based on the morphology seen in the color-magnitude
diagrams.  Their ``stream-like'' fields resemble the populations found
in the Giant Stream, while their ``disk-like'' fields reveal prominent
internmediate-age and recent star formation.  The Northern Spur
belongs to their ``disk-like'' fields, just as the G1 Clump and
Andromeda NE.  

Our data provide further support for the presence of intermediate-age
populations, since we found 3--4 PNe in the area of the G1 Clump and
7--10 PNe in the area of the Northern Spur. Our
Figures~\ref{fig:Sel_pos1} and \ref{fig:Obs_final} show the high
density of detected PNe in the area of the Northern Spur as compared
to the neighboring regions, and Figures~\ref{fig:gen_vels} and
\ref{fig:gen_stream} show that three of these PNe are kinematically
different and belong to the ``continuation of the Giant Stream''
sample.  In this scenario the progenitor of the Giant Stream
passed over, near, or through the Northern Spur area and lost there a
sufficiently large part of its mass to account for the PNe.
Considering the detected PNe, the total mass of the Northern Spur may
be estimated to be $6\times10^8 M_{\sun}$, of which 20--30\% could
have been contributed by the Giant Stream. Since the resulting mixture
of populations spread over a sufficiently large area and the resulting
range of the metallicities is unknown the detection of such a
proposed accreted component would be difficult observationally.
However, it would be interesting to compare abundances of different PNe
from the Northern Spur area with those from an in-situ population in
the same area and with PNe from Anromeda NE.

%---------------------------------------------------------------
\section{Summary and Conclusions}
\label{txt:summary}
%---------------------------------------------------------------

In this paper we present a method to identify extragalactic PNe based
on $ugri$ SDSS photometry and results from follow-up studies.  Our
results and conclusions can be summarized as follows:

1. We have developed a method to identify PN candidates in imaging
data of the SDSS using their unique characteristics in $ugri$
photometric data.  We apply and test this technique using \MA\ and its
large number of PNe.  Altogether, we identify 167 PN candidates in the
\MA\ area.

2. We demonstrate that our color-selection method for PN candidates
using SDSS $ugri$ filters can work very well for point-like sources at
distances of 90--800~kpc. The probability for the selected sources
to be contaminated with other types of objects is very low.  For
extended sources the probability that the selected candidates could be
contaminated by ELGs with strong emission lines and with redshift
$\le$ 0.1 is also very low.  But this contamination will surely grow
when the $(u-g)_0$ color criterion is relaxed:  $(u - g)_0 =
0\fm6-1\fm0$ is approximately the limit at which ELGs with strong
lines and PNe start to be comparable in SDSS colors.

3. We obtained spectroscopic follow-up observations of 80 PN
candidates in the \MA\ area.  These observations, additional
cross-identification work with other samples, and visual checks with
narrow-band images show that the efficiency of our method is at least
92\% for the area of our ``first priority'' candidates.  The
efficiency for the area of our ``second priority'' candidates drops
from $\sim100\%$ for magnitudes of $g<20\fm0$ to $\sim30\%$ for
magnitudes of $g>21\fm6$.

4. In general, the distribution of PNe in the outer region of \MA,
i.e., $8<R<20$~kpc along the minor axis follows the rotationally
supported low surface-brightness structure with an exponential scale
length of 3.21$\pm$0.14~kpc, the so-called extended disk suggested by
\citet{W05}.  This disk-like component is also visible in photometric
data from the Isaac Newton Telescope Wide Field Camera survey of \MA\
\citep[][]{Ibata05,Ibata07}.  We estimate the total stellar mass of
this structure to be $\sim10^{10} M_{\sun}$, which is equivalent of
the mass of M\,33.

5. Our spectroscopy confirms that we have found two new PNe in the
area of Andromeda NE \citep{Zucker04a}, a number consistent with the
number of PNe that can be expected in a stellar structure of this low
a luminosity ($\sim 5\times10^6 L_\odot$).  These two PNe are located
at projected distances of $\sim$46 Kpc and $\sim$40 Kpc along the
major axis from the center of \MA.

6. With the new PN data at hand we see a possible kinematic connection
between the Giant Stream and PNe in Andromeda\,NE suggesting that
Andromeda\,NE could be the core or remnant of the Giant Stream.  Using
the PN data we estimate the total mass of the Giant Stream progenitor
to be $\approx10^9 M_{\sun}$, which would imply that about 90\% of
stars were lost during the interaction with \MA.

6. Our data show an obvious kinematic connection between the
continuation of the Giant Stream and the Northern Spur.  We suggest
that 20 -- 30\% of stars in the Northern Spur area could belong to the
Giant Stream.

\acknowledgments

The Sloan Digital Sky Survey (SDSS) is a joint project of The
University of Chicago, Fermilab, the Institute for Advanced Study, the
Japan Participation Group, The Johns Hopkins University, the
Max-Planck-Institute for Astronomy (MPIA), the Max-Planck-Institute
for Astrophysics (MPA), New Mexico State University, Princeton
University, the United States Naval Observatory, and the University of
Washington. Apache Point Observatory, site of the SDSS telescopes, is
operated by the Astrophysical Research Consortium (ARC).

Funding for the project has been provided by the Alfred P.\ Sloan
Foundation, the SDSS member institutions, the National Aeronautics and
Space Administration, the National Science Foundation, the U.S.\
Department of Energy, the Japanese Monbukagakusho, and the Max Planck
Society. The SDSS Web site is http://www.sdss.org/.

A.Y.K acknowledges the support
from the National Research Foundation (NRF) of South Africa
and support from the Collaborative Research Area
``The Milky Way System'' (SFB 881) of the German Research Foundation
(DFG) during his visits to Heidelberg.
EKG acknowledges
support from the Swiss National Science Foundation through the grants
200021-101924/1 and 200020-105260/1.

\clearpage

%*********************************************************************
%*********************************************************************

%*********************************************************************
%*********************************************************************

%*********************************************************************
%*********************************************************************

%*********************************************************************
%*********************************************************************

%*********************************************************************
%*********************************************************************

%*********************************************************************
%*********************************************************************

%*********************************************************************
%*********************************************************************

%*********************************************************************
%*********************************************************************

%*********************************************************************
%*********************************************************************

%*********************************************************************
%*********************************************************************

%*********************************************************************
%*********************************************************************

%*********************************************************************
%*********************************************************************

%*********************************************************************
%*********************************************************************

%*********************************************************************
%*********************************************************************

%*********************************************************************
%*********************************************************************

%*********************************************************************
%*********************************************************************

%*********************************************************************
%*********************************************************************

\clearpage

\newcounter{tab}
\setcounter{tab}{0}
\newcommand{\numb}{\addtocounter{tab}{1}\arabic{tab}}

%
%...The general SDSS parameters of genuine PNe
%
\begin{deluxetable}{rllllllllccll}
\tablecolumns{13}
\rotate
%\tabletypesize{\tiny}
\tabletypesize{\footnotesize}
\tablecaption{SDSS positions and magnitudes of selected PN candidates in 
SDSS \MA\ data
\label{tbl-1}}
\tablewidth{0pt}
\tablehead{%
\colhead{\#} &\colhead{SDSS name} & \colhead{$\alpha(2000.0)$} & \colhead{$\delta(2000.0)$} &
\colhead{$u_0$} & \colhead{$g_0$} & \colhead{$r_0$} & \colhead{$i_0$} & \colhead{$z_0$} &
\colhead{pr$^a$} & \colhead{obs$^b$} & \colhead{Host$^c$} & \colhead{sample$^d$} \\
& & & & \colhead{mag} & \colhead{mag} & \colhead{mag} & \colhead{mag} & \colhead{mag} \\
&\colhead{(1)} & \colhead{(2)} & \colhead{(3)} & \colhead{(4)} &  \colhead{(5)}&
\colhead{(6)}  & \colhead{(7)} & \colhead{(8)} & \colhead{(9)} & \colhead{(10)} & \colhead{(11)} & 
\colhead{(12)}
}
\startdata
\qq & J001934+351248 & 00:19:34.61 & +35:12:48.60 & 22.74$\pm$0.41 & 22.09$\pm$0.09 & 22.80$\pm$0.18 & 23.27$\pm$0.38  & 22.22$\pm$0.60 &  2  & 0  &      &                \\
\qq & J003206+375301 & 00:32:06.12 & +37:53:01.68 & 22.83$\pm$0.50 & 22.07$\pm$0.10 & 22.54$\pm$0.17 & 22.96$\pm$0.38  & 22.84$\pm$0.57 &  2  & 0  &      &                \\
\qq & J003338+395251 & 00:33:38.81 & +39:52:51.60 & 20.69$\pm$0.07 & 19.99$\pm$0.03 & 20.44$\pm$0.23 & 24.23$\pm$0.60  & 19.76$\pm$0.10 &  2  & 0  &      &                \\
\qq & J003351+402941 & 00:33:51.10 & +40:29:41.28 & 22.82$\pm$0.47 & 20.30$\pm$0.03 & 21.49$\pm$0.06 & 22.94$\pm$0.30  & 23.00$\pm$0.54 &  1  & 0  & SW   & NF87           \\
\qq & J003419+404041 & 00:34:19.82 & +40:40:41.16 & 21.72$\pm$0.19 & 21.07$\pm$0.07 & 21.60$\pm$0.10 & 22.36$\pm$0.31  & 22.89$\pm$0.57 &  2  & 0  &      &                \\
\qq & J003447+393612 & 00:34:47.57 & +39:36:12.24 & 22.05$\pm$0.22 & 20.55$\pm$0.03 & 21.48$\pm$0.06 & 22.95$\pm$0.28  & 22.79$\pm$0.61 &  1  & 1  & G1   &                \\
\qq & J003451+395429 & 00:34:51.19 & +39:54:29.16 & 22.29$\pm$0.37 & 20.64$\pm$0.04 & 21.44$\pm$0.08 & 23.02$\pm$0.45  & 22.30$\pm$0.66 &  1  & 1  & G1   &                \\
\qq & J003451+402254 & 00:34:51.79 & +40:22:54.12 & 21.89$\pm$0.17 & 20.17$\pm$0.02 & 21.25$\pm$0.05 & 22.42$\pm$0.20  & 22.56$\pm$0.51 &  1  & 1  & SW   &                \\
\qq & J003509+392824 & 00:35:09.14 & +39:28:24.96 & 23.08$\pm$0.48 & 21.54$\pm$0.06 & 22.52$\pm$0.20 & 23.76$\pm$0.56  & 22.69$\pm$0.60 &  1  & 0  & G1   & NF87,M06       \\
\qq & J003533+405521 & 00:35:33.41 & +40:55:21.00 & 21.78$\pm$0.22 & 20.82$\pm$0.04 & 21.43$\pm$0.06 & 22.85$\pm$0.30  & 22.00$\pm$0.44 &  2  & 1  & SW   & M06            \\
\qq & J003548+404039 & 00:35:48.98 & +40:40:39.00 & 23.12$\pm$0.72 & 21.64$\pm$0.06 & 22.06$\pm$0.16 & 22.73$\pm$0.60  & 22.77$\pm$0.83 &  2  & 0  &      &                \\
\qq & J003555+403126 & 00:35:55.70 & +40:31:26.04 & 24.28$\pm$0.74 & 21.30$\pm$0.05 & 22.28$\pm$0.11 & 23.15$\pm$0.36  & 23.60$\pm$0.38 &  1  & 0  & SW   & NF87           \\
\qq & J003559+414050 & 00:35:59.83 & +41:40:50.52 & 22.95$\pm$0.58 & 21.83$\pm$0.07 & 22.58$\pm$0.20 & 23.20$\pm$0.52  & 22.40$\pm$0.65 &  2  & 1  &      &                \\
\qq & J003559+405033 & 00:35:59.98 & +40:50:33.36 & 22.27$\pm$0.38 & 21.09$\pm$0.04 & 21.82$\pm$0.11 & 22.81$\pm$0.46  & 22.18$\pm$1.19 &  1  & 0  &      & NF87,M06       \\
\qq & J003600+392932 & 00:36:00.12 & +39:29:32.64 & 22.08$\pm$0.26 & 20.40$\pm$0.03 & 21.65$\pm$0.11 & 23.31$\pm$0.60  & 21.80$\pm$0.59 &  1  & 0  & GL   & NF87           \\
\qq & J003605+403021 & 00:36:05.50 & +40:30:21.60 & 22.61$\pm$0.28 & 21.45$\pm$0.05 & 22.20$\pm$0.11 & 24.17$\pm$0.54  & 22.29$\pm$0.48 &  1  & 0  & SW   & NF87           \\
\qq & J003612+393541 & 00:36:12.60 & +39:35:41.64 & 22.12$\pm$0.27 & 20.89$\pm$0.04 & 21.75$\pm$0.11 & 22.22$\pm$0.28  & 23.78$\pm$0.81 &  1  & 1  & GL   & M06            \\
\qq & J003619+413807 & 00:36:19.08 & +41:38:07.44 & 20.44$\pm$0.07 & 19.84$\pm$0.02 & 20.25$\pm$0.14 & 24.22$\pm$0.75  & 19.32$\pm$0.07 &  2  & 0  &      &                \\
\qq & J003628+393526 & 00:36:28.78 & +39:35:26.16 & 22.62$\pm$0.40 & 20.19$\pm$0.03 & 21.27$\pm$0.07 & 22.87$\pm$0.38  & 23.49$\pm$0.60 &  1  & 1  & GL   & M06            \\
\qq & J003634+395038 & 00:36:34.46 & +39:50:38.40 & 22.67$\pm$0.37 & 21.91$\pm$0.08 & 22.38$\pm$0.12 & 23.29$\pm$0.34  & 22.08$\pm$0.51 &  2  & 0  &      &                \\
\qq & J003635+401737 & 00:36:35.33 & +40:17:37.68 & 22.54$\pm$0.30 & 21.65$\pm$0.06 & 22.46$\pm$0.20 & 24.19$\pm$0.76  & 23.80$\pm$0.60 &  2  & 1  &      &                \\
\qq & J003644+400354 & 00:36:44.74 & +40:03:54.36 & 22.74$\pm$0.40 & 21.94$\pm$0.07 & 22.39$\pm$0.13 & 22.89$\pm$0.28  & 22.75$\pm$0.67 &  2  & 0  &      &                \\
\qq & J003653+401331 & 00:36:53.90 & +40:13:31.44 & 22.92$\pm$0.65 & 21.72$\pm$0.07 & 22.21$\pm$0.16 & 22.74$\pm$0.43  & 21.96$\pm$0.80 &  2  & 3  &      &                \\
\qq & J003654+395911 & 00:36:54.94 & +39:59:11.04 & 23.01$\pm$0.48 & 22.10$\pm$0.08 & 22.53$\pm$0.14 & 22.76$\pm$0.23  & 22.71$\pm$0.63 &  2  & 0  &      &                \\
\qq & J003657+395436 & 00:36:57.05 & +39:54:36.00 & 22.58$\pm$0.35 & 21.96$\pm$0.07 & 22.46$\pm$0.13 & 22.81$\pm$0.26  & 23.13$\pm$0.58 &  2  & 0  &      &                \\
\qq & J003703+394445 & 00:37:03.48 & +39:44:45.24 & 23.41$\pm$0.74 & 22.01$\pm$0.10 & 22.44$\pm$0.17 & 23.35$\pm$0.74  & 22.62$\pm$0.81 &  2  & 0  &      &                \\
\qq & J003707+383856 & 00:37:07.13 & +38:38:56.76 & 22.26$\pm$0.27 & 21.25$\pm$0.04 & 22.25$\pm$0.12 & 23.29$\pm$0.34  & 21.82$\pm$0.35 &  1  & 1  &      &                \\
\qq & J003717+402437 & 00:37:17.30 & +40:24:37.80 & 23.09$\pm$0.79 & 20.69$\pm$0.03 & 21.31$\pm$0.08 & 22.57$\pm$0.33  & 22.62$\pm$0.87 &  1  & 1  &      &                \\
\qq & J003721+395050 & 00:37:21.12 & +39:50:50.64 & 22.54$\pm$0.38 & 20.73$\pm$0.04 & 21.51$\pm$0.08 & 22.81$\pm$0.42  & 22.25$\pm$0.79 &  1  & 0  &      & NF87           \\
\qq & J003721+404142 & 00:37:21.82 & +40:41:42.72 & 22.80$\pm$0.39 & 21.81$\pm$0.07 & 22.69$\pm$0.16 & 23.24$\pm$0.39  & 22.86$\pm$0.49 &  2  & 1  &      & M06            \\
\qq & J003726+401724 & 00:37:26.16 & +40:17:24.72 & 22.29$\pm$0.29 & 21.57$\pm$0.05 & 22.06$\pm$0.10 & 22.31$\pm$0.22  & 21.99$\pm$0.52 &  2  & 4  &      &                \\
\qq & J003734+402302 & 00:37:34.13 & +40:23:02.76 & 21.82$\pm$0.29 & 20.93$\pm$0.04 & 21.49$\pm$0.09 & 22.10$\pm$0.22  & 20.84$\pm$0.30 &  2  & 4  &      &                \\
\qq & J003736+402802 & 00:37:36.17 & +40:28:02.28 & 22.37$\pm$0.34 & 20.69$\pm$0.03 & 21.33$\pm$0.08 & 23.01$\pm$0.44  & 21.87$\pm$0.66 &  1  & 1  &      &                \\
\qq & J003736+394707 & 00:37:36.34 & +39:47:07.44 & 24.50$\pm$0.97 & 21.79$\pm$0.15 & 22.93$\pm$0.63 & 24.24$\pm$0.80  & 20.45$\pm$0.27 &  2  & 3  &      &                \\
\qq & J003736+393935 & 00:37:36.58 & +39:39:35.64 & 23.67$\pm$0.58 & 21.34$\pm$0.04 & 22.74$\pm$0.18 & 23.70$\pm$0.45  & 23.02$\pm$0.48 &  1  & 0  &      &                \\
\qq & J003738+395003 & 00:37:38.81 & +39:50:03.48 & 21.95$\pm$0.18 & 20.54$\pm$0.03 & 21.33$\pm$0.07 & 22.36$\pm$0.28  & 23.29$\pm$0.78 &  1  & 0  &      & NF87           \\
\qq & J003739+393009 & 00:37:39.94 & +39:30:09.36 & 23.06$\pm$0.38 & 21.80$\pm$0.06 & 22.62$\pm$0.15 & 22.99$\pm$0.28  & 22.91$\pm$0.50 &  2  & 1  & GL   &                \\
\qq & J003742+395235 & 00:37:42.98 & +39:52:35.76 & 23.38$\pm$0.72 & 20.91$\pm$0.04 & 21.68$\pm$0.10 & 22.79$\pm$0.40  & 21.92$\pm$0.69 &  1  & 0  &      & NF87           \\
\qq & J003745+403127 & 00:37:45.65 & +40:31:27.48 & 23.35$\pm$0.89 & 21.84$\pm$0.08 & 22.38$\pm$0.20 & 22.74$\pm$0.48  & 23.02$\pm$0.83 &  2  & 3  &      &                \\
\qq & J003749+405354 & 00:37:49.70 & +40:53:54.24 & 21.47$\pm$0.13 & 20.53$\pm$0.03 & 21.11$\pm$0.04 & 22.21$\pm$0.19  & 21.81$\pm$0.39 &  2  & 1  &      & M06            \\
\qq & J003751+404547 & 00:37:51.29 & +40:45:47.88 & 22.19$\pm$0.35 & 20.42$\pm$0.03 & 21.16$\pm$0.07 & 22.72$\pm$0.39  & 23.69$\pm$0.54 &  1  & 1  &      & M06            \\
\qq & J003808+395645 & 00:38:08.38 & +39:56:45.60 & 20.55$\pm$0.08 & 19.94$\pm$0.04 & 20.37$\pm$0.07 & 21.30$\pm$0.20  & 21.43$\pm$0.60 &  2  & 4  &      &                \\
\qq & J003818+400634 & 00:38:18.53 & +40:06:34.92 & 24.18$\pm$1.17 & 22.03$\pm$0.10 & 23.05$\pm$0.53 & 23.46$\pm$0.72  & 22.74$\pm$0.76 &  2  & 4  &      &                \\
\qq & J003818+421252 & 00:38:18.58 & +42:12:52.20 & 22.93$\pm$0.58 & 21.69$\pm$0.07 & 22.15$\pm$0.13 & 22.56$\pm$0.30  & 22.44$\pm$1.44 &  2  & 0  &      &                \\
\qq & J003827+404711 & 00:38:27.14 & +40:47:11.40 & 23.41$\pm$0.87 & 21.89$\pm$0.09 & 22.37$\pm$0.20 & 22.93$\pm$0.48  & 22.17$\pm$0.69 &  2  & 0  &      &                \\
\qq & J003832+401310 & 00:38:32.81 & +40:13:10.20 & 21.78$\pm$0.23 & 21.02$\pm$0.04 & 21.47$\pm$0.09 & 22.06$\pm$0.20  & 22.08$\pm$0.63 &  2  & 4  &      &                \\
\qq & J003841+394739 & 00:38:41.30 & +39:47:39.84 & 23.43$\pm$0.53 & 22.12$\pm$0.08 & 22.80$\pm$0.19 & 24.40$\pm$0.48  & 22.28$\pm$0.53 &  2  & 1  & GL   & HK04           \\
\qq & J003846+412819 & 00:38:46.15 & +41:28:19.20 & 22.43$\pm$0.32 & 20.36$\pm$0.03 & 21.33$\pm$0.07 & 23.20$\pm$0.47  & 22.09$\pm$0.52 &  1  & 1  & LO   & M06            \\
\qq & J003848+413937 & 00:38:48.38 & +41:39:37.44 & 23.02$\pm$0.49 & 22.19$\pm$0.10 & 22.89$\pm$0.26 & 23.94$\pm$0.70  & 21.66$\pm$0.42 &  2  & 1  & LO   & M06            \\
\qq & J003849+400153 & 00:38:49.85 & +40:01:53.40 & 23.00$\pm$0.65 & 21.84$\pm$0.08 & 24.00$\pm$0.65 & 25.94$\pm$0.30  & 24.01$\pm$0.44 &  2  & 4  &      &                \\
\qq & J003849+400544 & 00:38:49.85 & +40:05:44.88 & 22.64$\pm$0.32 & 21.31$\pm$0.06 & 21.93$\pm$0.17 & 22.36$\pm$0.29  & 21.50$\pm$0.48 &  1  & 4  &      &                \\
\qq & J003852+404130 & 00:38:52.42 & +40:41:30.48 & 22.70$\pm$0.93 & 22.05$\pm$0.10 & 22.46$\pm$0.25 & 24.38$\pm$0.78  & 22.27$\pm$0.70 &  2  & 0  &      &                \\
\qq & J003855+410655 & 00:38:55.03 & +41:06:55.08 & 22.80$\pm$0.40 & 20.54$\pm$0.03 & 21.73$\pm$0.08 & 22.47$\pm$0.24  & 22.30$\pm$0.47 &  1  & 1  &      & M06            \\
\qq & J003901+415111 & 00:39:01.08 & +41:51:11.16 & 22.02$\pm$1.05 & 20.56$\pm$0.03 & 21.55$\pm$0.07 & 22.78$\pm$0.30  & 21.25$\pm$0.28 &  1  & 1  & LO   & M06            \\
\qq & J003902+402250 & 00:39:02.52 & +40:22:50.16 & 23.46$\pm$0.91 & 20.75$\pm$0.04 & 21.76$\pm$0.13 & 25.14$\pm$0.60  & 21.88$\pm$0.60 &  1  & 1  &      & HK04,M06       \\
\qq & J003903+394558 & 00:39:03.38 & +39:45:58.32 & 22.74$\pm$0.33 & 20.58$\pm$0.03 & 21.36$\pm$0.05 & 21.81$\pm$0.11  & 21.94$\pm$0.42 &  1  & 0  & GL   & NF87,HK04      \\
\qq & J003903+395329 & 00:39:03.77 & +39:53:29.04 & 23.22$\pm$0.47 & 21.36$\pm$0.05 & 22.54$\pm$0.16 & 23.61$\pm$0.47  & 22.99$\pm$0.53 &  1  & 2  &      & HK04           \\
\qq & J003906+401459 & 00:39:06.62 & +40:14:59.64 & 22.89$\pm$0.58 & 20.44$\pm$0.03 & 21.26$\pm$0.08 & 22.33$\pm$0.30  & 21.40$\pm$0.42 &  1  & 0  &      & NF87,HK04,M06  \\
\qq & J003909+401120 & 00:39:09.31 & +40:11:20.76 & 23.74$\pm$1.26 & 20.55$\pm$0.03 & 21.39$\pm$0.09 & 22.02$\pm$0.24  & 22.20$\pm$0.72 &  1  & 0  &      & NF87,HK04,M06  \\
\qq & J003915+404811 & 00:39:15.91 & +40:48:11.52 & 22.15$\pm$0.47 & 21.36$\pm$0.09 & 22.10$\pm$0.25 & 22.51$\pm$0.38  & 21.68$\pm$0.60 &  2  & 4  &      &                \\
\qq & J003916+402615 & 00:39:16.73 & +40:26:15.72 & 22.74$\pm$0.77 & 21.00$\pm$0.09 & 21.45$\pm$0.13 & 21.77$\pm$0.34  & 21.60$\pm$0.34 &  1  & 0  &      & M06            \\
\qq & J003918+402126 & 00:39:18.07 & +40:21:26.64 & 21.38$\pm$0.22 & 20.59$\pm$0.04 & 21.05$\pm$0.08 & 21.52$\pm$0.18  & 22.27$\pm$0.71 &  2  & 4  &      &                \\
\qq & J003918+400919 & 00:39:18.55 & +40:09:19.80 & 21.96$\pm$0.27 & 20.58$\pm$0.03 & 22.20$\pm$0.18 & 22.52$\pm$0.55  & 23.29$\pm$0.68 &  1  & 0  &      & NF87,M06,HK04  \\
\qq & J003922+410657 & 00:39:22.58 & +41:06:57.24 & 23.26$\pm$0.74 & 20.42$\pm$0.03 & 21.43$\pm$0.08 & 22.87$\pm$0.56  & 23.19$\pm$0.72 &  1  & 1  &      & M06            \\
\qq & J003924+400702 & 00:39:24.53 & +40:07:02.28 & 22.50$\pm$0.29 & 20.58$\pm$0.03 & 21.62$\pm$0.07 & 23.69$\pm$0.46  & 22.33$\pm$1.44 &  1  & 0  &      & NF87,HK04,M06  \\
\qq & J003929+405201 & 00:39:29.09 & +40:52:01.92 & 21.70$\pm$0.24 & 21.04$\pm$0.04 & 21.74$\pm$0.13 & 24.64$\pm$0.68  & 21.76$\pm$0.60 &  2  & 0  &      & M06            \\
\qq & J003936+410608 & 00:39:36.17 & +41:06:08.64 & 22.67$\pm$0.96 & 21.65$\pm$0.07 & 22.18$\pm$0.16 & 22.39$\pm$0.27  & 22.36$\pm$0.76 &  2  & 0  &      &                \\
\qq & J003940+402554 & 00:39:40.44 & +40:25:54.84 & 22.06$\pm$0.37 & 20.99$\pm$0.06 & 21.45$\pm$0.12 & 22.71$\pm$0.36  & 22.02$\pm$0.58 &  1  & 4  &      &                \\
\qq & J003941+410145 & 00:39:41.81 & +41:01:45.84 & 22.83$\pm$0.56 & 22.04$\pm$0.10 & 22.55$\pm$0.26 & 23.09$\pm$0.38  & 22.32$\pm$0.75 &  2  & 0  &      &                \\
\qq & J003942+410221 & 00:39:42.53 & +41:02:21.12 & 23.13$\pm$1.11 & 21.51$\pm$0.06 & 22.44$\pm$0.21 & 24.03$\pm$0.74  & 22.52$\pm$0.78 &  1  & 0  &      & M06            \\
\qq & J003945+403142 & 00:39:45.38 & +40:31:42.60 & 21.87$\pm$0.36 & 20.17$\pm$0.03 & 21.19$\pm$0.12 & 23.23$\pm$0.73  & 21.89$\pm$0.56 &  1  & 1  &      & HK04,M06       \\
\qq & J003946+393842 & 00:39:46.70 & +39:38:42.72 & 21.81$\pm$0.24 & 21.13$\pm$0.04 & 21.64$\pm$0.10 & 23.87$\pm$0.73  & 21.97$\pm$0.66 &  2  & 0  &      & HK04           \\
\qq & J003948+410841 & 00:39:48.19 & +41:08:41.64 & 22.67$\pm$0.53 & 20.92$\pm$0.12 & 21.72$\pm$0.21 & 23.33$\pm$0.71  & 22.08$\pm$0.38 &  1  & 0  &      &                \\
\qq & J003948+411043 & 00:39:48.77 & +41:10:43.32 & 22.75$\pm$0.52 & 21.19$\pm$0.05 & 22.30$\pm$0.18 & 23.83$\pm$0.68  & 21.76$\pm$0.57 &  1  & 1  &      &                \\
\qq & J003951+402327 & 00:39:51.86 & +40:23:27.96 & 20.59$\pm$0.13 & 19.85$\pm$0.03 & 20.26$\pm$0.04 & 20.51$\pm$0.08  & 20.77$\pm$0.26 &  2  & 4  &      &                \\
\qq & J003954+402338 & 00:39:54.17 & +40:23:38.40 & 21.43$\pm$0.26 & 20.22$\pm$0.03 & 21.14$\pm$0.12 & 21.51$\pm$0.31  & 20.72$\pm$0.25 &  1  & 4  &      &                \\
\qq & J003954+395103 & 00:39:54.38 & +39:51:03.24 & 23.04$\pm$0.82 & 21.05$\pm$0.04 & 22.55$\pm$0.19 & 24.09$\pm$0.70  & 21.92$\pm$0.57 &  1  & 0  & GL   & NF87           \\
\qq & J003956+414306 & 00:39:56.88 & +41:43:06.96 & 24.01$\pm$0.85 & 21.59$\pm$0.07 & 23.06$\pm$0.31 & 23.33$\pm$0.55  & 23.04$\pm$0.63 &  1  & 0  &NGC205& M06            \\
\qq & J003956+410507 & 00:39:56.98 & +41:05:07.44 & 21.95$\pm$0.29 & 20.41$\pm$0.03 & 21.08$\pm$0.07 & 23.04$\pm$0.35  & 22.84$\pm$0.78 &  1  & 1  &      & M06            \\
\qq & J003958+402450 & 00:39:58.97 & +40:24:50.40 & 20.39$\pm$0.21 & 19.53$\pm$0.04 & 19.96$\pm$0.13 & 20.48$\pm$0.14  & 21.21$\pm$0.38 &  2  & 4  &      &                \\
\qq & J004000+410409 & 00:40:00.86 & +41:04:09.12 & 22.30$\pm$1.00 & 20.26$\pm$0.03 & 21.12$\pm$0.07 & 21.89$\pm$0.19  & 21.46$\pm$0.38 &  1  & 1  &      & M06            \\
\qq & J004003+402721 & 00:40:03.02 & +40:27:21.60 & 20.84$\pm$0.15 & 19.60$\pm$0.02 & 20.03$\pm$0.04 & 20.28$\pm$0.08  & 20.51$\pm$0.20 &  2  & 4  &      &                \\
\qq & J004012+401302 & 00:40:12.79 & +40:13:02.28 & 22.97$\pm$0.44 & 21.36$\pm$0.05 & 22.92$\pm$0.22 & 24.08$\pm$0.49  & 22.48$\pm$0.50 &  1  & 0  &      & HK04,M06       \\
\qq & J004013+401026 & 00:40:13.97 & +40:10:26.04 & 22.27$\pm$0.25 & 20.50$\pm$0.03 & 21.45$\pm$0.06 & 22.37$\pm$0.18  & 21.90$\pm$0.37 &  1  & 0  &      & NF87,HK04      \\
\qq & J004035+421031 & 00:40:35.35 & +42:10:31.80 & 22.16$\pm$0.30 & 21.24$\pm$0.05 & 21.78$\pm$0.08 & 23.17$\pm$0.36  & 21.99$\pm$0.48 &  2  & 1  & LO   & M06            \\
\qq & J004043+395705 & 00:40:43.54 & +39:57:05.40 & 22.79$\pm$0.66 & 21.11$\pm$0.04 & 22.22$\pm$0.17 & 25.49$\pm$0.43  & 23.66$\pm$0.76 &  1  & 0  &      & NF87           \\
\qq & J004056+402008 & 00:40:56.88 & +40:20:08.88 & 21.98$\pm$0.20 & 20.12$\pm$0.02 & 21.23$\pm$0.05 & 22.32$\pm$0.27  & 21.54$\pm$0.33 &  1  & 1  &      & HK04           \\
\qq & J004119+415146 & 00:41:19.08 & +41:51:46.44 & 22.97$\pm$0.43 & 22.01$\pm$0.08 & 22.76$\pm$0.19 & 24.19$\pm$0.56  & 23.41$\pm$0.44 &  2  & 2  &      & M06            \\
\qq & J004133+412303 & 00:41:33.43 & +41:23:03.12 & 23.87$\pm$1.01 & 21.67$\pm$0.10 & 22.16$\pm$0.16 & 24.40$\pm$0.69  & 22.27$\pm$0.80 &  2  & 0  &      &                \\
\qq & J004133+412258 & 00:41:33.91 & +41:22:58.08 & 22.23$\pm$0.96 & 20.91$\pm$0.06 & 21.64$\pm$0.12 & 22.25$\pm$0.15  & 21.09$\pm$0.40 &  1  & 0  &      &                \\
\qq & J004138+395459 & 00:41:38.18 & +39:54:59.76 & 22.35$\pm$0.33 & 21.29$\pm$0.04 & 21.82$\pm$0.09 & 23.37$\pm$0.37  & 22.31$\pm$0.47 &  1  & 0  &      & NF87,HK04,M06  \\
\qq & J004212+423134 & 00:42:12.43 & +42:31:34.32 & 22.75$\pm$0.53 & 21.54$\pm$0.06 & 22.74$\pm$0.19 & 23.11$\pm$0.40  & 22.96$\pm$0.56 &  1  & 1  & LO   &                \\
\qq & J004216+423826 & 00:42:16.51 & +42:38:26.16 & 24.68$\pm$0.96 & 22.03$\pm$0.09 & 22.44$\pm$0.16 & 22.98$\pm$0.38  & 23.46$\pm$0.47 &  2  & 0  &      &                \\
\qq & J004222+414313 & 00:42:22.15 & +41:43:13.08 & 23.23$\pm$0.76 & 21.90$\pm$0.09 & 22.85$\pm$0.28 & 24.21$\pm$0.70  & 22.41$\pm$0.84 &  2  & 0  &      &                \\
\qq & J004240+413548 & 00:42:40.61 & +41:35:48.48 & 22.13$\pm$1.09 & 20.31$\pm$0.05 & 21.30$\pm$0.13 & 24.24$\pm$0.63  & 23.52$\pm$0.54 &  1  & 0  &      & M06            \\
\qq & J004244+401733 & 00:42:44.66 & +40:17:33.72 & 22.61$\pm$0.48 & 21.61$\pm$0.06 & 22.39$\pm$0.20 & 23.21$\pm$0.52  & 22.29$\pm$0.58 &  2  & 3  &      &                \\
\qq & J004252+402903 & 00:42:52.10 & +40:29:03.84 & 22.27$\pm$0.34 & 20.98$\pm$0.04 & 22.12$\pm$0.16 & 23.60$\pm$0.61  & 22.69$\pm$0.64 &  1  & 1  &      & HK04,M06       \\
\qq & J004258+424732 & 00:42:58.46 & +42:47:32.64 & 22.18$\pm$0.33 & 21.09$\pm$0.04 & 21.72$\pm$0.09 & 23.63$\pm$0.56  & 22.68$\pm$0.57 &  1  & 1  & LO   &                \\
\qq & J004258+401140 & 00:42:58.90 & +40:11:40.20 & 22.15$\pm$0.27 & 21.53$\pm$0.05 & 21.94$\pm$0.10 & 22.54$\pm$0.23  & 21.94$\pm$0.42 &  2  & 0  &      &                \\
\qq & J004259+420213 & 00:42:59.09 & +42:02:13.92 & 22.32$\pm$1.02 & 20.55$\pm$0.03 & 21.46$\pm$0.06 & 22.75$\pm$0.34  & 22.86$\pm$0.53 &  1  & 1  &      & M06            \\
\qq & J004303+401949 & 00:43:03.38 & +40:19:49.44 & 22.49$\pm$0.37 & 21.72$\pm$0.06 & 22.15$\pm$0.12 & 22.73$\pm$0.25  & 22.07$\pm$0.43 &  2  & 0  &      &                \\
\qq & J004306+404041 & 00:43:06.19 & +40:40:41.88 & 23.77$\pm$2.68 & 21.79$\pm$0.08 & 22.40$\pm$0.21 & 22.61$\pm$0.31  & 23.25$\pm$0.56 &  2  & 0  &      &                \\
\qq & J004311+422045 & 00:43:11.18 & +42:20:45.60 & 21.72$\pm$0.18 & 20.45$\pm$0.03 & 21.59$\pm$0.07 & 23.01$\pm$0.38  & 22.26$\pm$0.50 &  1  & 1  &      & M06            \\
\qq & J004327+424203 & 00:43:27.24 & +42:42:03.60 & 22.28$\pm$0.33 & 21.12$\pm$0.05 & 22.15$\pm$0.14 & 22.85$\pm$0.47  & 21.80$\pm$0.47 &  1  & 1  & LO   &                \\
\qq & J004328+415155 & 00:43:28.46 & +41:51:55.44 & 23.60$\pm$1.07 & 21.69$\pm$0.09 & 22.33$\pm$0.19 & 22.55$\pm$0.48  & 23.51$\pm$0.66 &  2  & 0  &      & M06            \\
\qq & J004332+415842 & 00:43:32.35 & +41:58:42.24 & 21.84$\pm$0.27 & 20.57$\pm$0.03 & 21.19$\pm$0.06 & 22.06$\pm$0.21  & 21.99$\pm$0.78 &  1  & 1  & GS   & M06            \\
\qq & J004336+414942 & 00:43:36.58 & +41:49:42.96 & 21.86$\pm$0.49 & 20.26$\pm$0.04 & 21.25$\pm$0.09 & 22.38$\pm$0.45  & 22.05$\pm$0.89 &  1  & 0  &      & M06            \\
\qq & J004342+422235 & 00:43:42.53 & +42:22:35.40 & 22.81$\pm$0.38 & 20.80$\pm$0.04 & 21.83$\pm$0.10 & 22.70$\pm$0.33  & 22.05$\pm$0.35 &  1  & 1  &      & M06            \\
\qq & J004352+421835 & 00:43:52.22 & +42:18:35.28 & 22.48$\pm$0.34 & 21.62$\pm$0.06 & 22.04$\pm$0.11 & 22.32$\pm$0.22  & 22.54$\pm$0.53 &  2  & 0  &      &                \\
\qq & J004402+421716 & 00:44:02.35 & +42:17:16.44 & 22.73$\pm$0.41 & 20.51$\pm$0.03 & 21.46$\pm$0.06 & 22.51$\pm$0.25  & 22.65$\pm$0.53 &  1  & 1  &      & M06            \\
\qq & J004403+422746 & 00:44:03.12 & +42:27:46.44 & 22.57$\pm$0.36 & 20.41$\pm$0.03 & 21.37$\pm$0.06 & 22.74$\pm$0.32  & 21.63$\pm$0.47 &  1  & 1  &      &                \\
\qq & J004410+412420 & 00:44:10.61 & +41:24:20.16 & 20.79$\pm$0.41 & 19.41$\pm$0.05 & 20.71$\pm$0.12 & 21.16$\pm$0.20  & 20.19$\pm$0.26 &  2  & 4  &      & M06            \\
\qq & J004421+422450 & 00:44:21.50 & +42:24:50.04 & 22.70$\pm$0.64 & 20.90$\pm$0.04 & 22.06$\pm$0.11 & 23.48$\pm$0.55  & 21.88$\pm$0.42 &  1  & 2  &      & M06            \\
\qq & J004430+420857 & 00:44:30.74 & +42:08:57.48 & 23.41$\pm$0.92 & 22.19$\pm$0.11 & 22.86$\pm$0.26 & 25.11$\pm$0.51  & 23.23$\pm$0.79 &  2  & 4  &      & M06            \\
\qq & J004450+420506 & 00:44:50.86 & +42:05:06.36 & 23.46$\pm$0.97 & 21.57$\pm$0.07 & 21.97$\pm$0.13 & 23.05$\pm$0.37  & 22.66$\pm$0.70 &  1  & 0  &      & M06            \\
\qq & J004452+421316 & 00:44:52.92 & +42:13:16.68 & 22.22$\pm$0.38 & 21.39$\pm$0.05 & 21.90$\pm$0.12 & 22.13$\pm$0.22  & 21.89$\pm$0.65 &  2  & 4  &      &                \\
\qq & J004454+421738 & 00:44:54.77 & +42:17:38.04 & 23.07$\pm$0.73 & 21.41$\pm$0.05 & 22.36$\pm$0.17 & 24.81$\pm$0.64  & 22.73$\pm$0.82 &  1  & 1  &      &                \\
\qq & J004501+421423 & 00:45:01.49 & +42:14:23.64 & 21.24$\pm$0.17 & 20.64$\pm$0.05 & 21.07$\pm$0.12 & 21.33$\pm$0.23  & 20.73$\pm$0.34 &  2  & 4  &      &                \\
\qq & J004523+421521 & 00:45:23.18 & +42:15:21.96 & 22.76$\pm$0.56 & 21.95$\pm$0.09 & 22.86$\pm$0.31 & 24.34$\pm$0.76  & 22.60$\pm$0.81 &  2  & 4  &      &                \\
\qq & J004537+422430 & 00:45:37.70 & +42:24:30.60 & 24.86$\pm$0.92 & 21.99$\pm$0.08 & 22.68$\pm$0.32 & 24.80$\pm$0.69  & 22.64$\pm$0.81 &  2  & 4  &      &                \\
\qq & J004542+425526 & 00:45:42.60 & +42:55:26.04 & 22.05$\pm$0.23 & 20.46$\pm$0.03 & 21.26$\pm$0.06 & 21.64$\pm$0.12  & 21.14$\pm$0.24 &  1  & 1  & SP   & NF87,M06       \\
\qq & J004544+422912 & 00:45:44.35 & +42:29:12.12 & 23.15$\pm$0.70 & 22.20$\pm$0.10 & 22.66$\pm$0.45 & 24.78$\pm$0.70  & 23.67$\pm$0.56 &  2  & 4  &      &                \\
\qq & J004552+423652 & 00:45:52.80 & +42:36:52.92 & 21.93$\pm$0.29 & 20.36$\pm$0.03 & 21.10$\pm$0.06 & 22.56$\pm$0.42  & 23.82$\pm$0.51 &  1  & 1  &      & NF87,M06       \\
\qq & J004613+424028 & 00:46:13.82 & +42:40:28.56 & 21.76$\pm$0.22 & 20.59$\pm$0.03 & 21.52$\pm$0.09 & 23.17$\pm$0.49  & 23.44$\pm$0.65 &  1  & 1  &      & NF87,M06       \\
\qq & J004619+412230 & 00:46:19.92 & +41:22:30.36 & 21.42$\pm$0.19 & 20.81$\pm$0.04 & 21.22$\pm$0.07 & 21.72$\pm$0.14  & 21.74$\pm$0.51 &  2  & 0  &      &                \\
\qq & J004626+430043 & 00:46:26.50 & +43:00:43.20 & 21.92$\pm$0.21 & 20.18$\pm$0.02 & 21.07$\pm$0.05 & 22.04$\pm$0.16  & 21.77$\pm$0.37 &  1  & 1  & GS   & NF87,M06       \\
\qq & J004641+435903 & 00:46:41.62 & +43:59:03.48 & 23.96$\pm$1.08 & 21.54$\pm$0.06 & 22.51$\pm$0.18 & 23.13$\pm$0.52  & 22.01$\pm$0.52 &  1  & 0  &      & JF86           \\
\qq & J004642+420835 & 00:46:42.94 & +42:08:35.52 & 21.71$\pm$0.29 & 19.98$\pm$0.03 & 20.99$\pm$0.07 & 21.92$\pm$0.34  & 22.08$\pm$0.62 &  1  & 1  &      & M06            \\
\qq & J004648+412352 & 00:46:48.07 & +41:23:52.80 & 22.59$\pm$0.50 & 21.91$\pm$0.10 & 22.52$\pm$0.20 & 23.78$\pm$0.63  & 22.49$\pm$0.73 &  2  & 0  &      &                \\
\qq & J004652+431058 & 00:46:52.15 & +43:10:58.80 & 22.64$\pm$0.37 & 20.96$\pm$0.04 & 22.10$\pm$0.11 & 23.86$\pm$0.59  & 22.34$\pm$0.50 &  1  & 1  & GS   &                \\
\qq & J004655+422401 & 00:46:55.46 & +42:24:01.80 & 21.65$\pm$0.21 & 19.99$\pm$0.02 & 21.01$\pm$0.05 & 21.55$\pm$0.11  & 20.39$\pm$0.17 &  1  & 1  &      & M06            \\
\qq & J004659+423758 & 00:46:59.28 & +42:37:58.08 & 22.18$\pm$0.32 & 20.51$\pm$0.03 & 21.33$\pm$0.08 & 22.92$\pm$0.42  & 22.88$\pm$0.79 &  1  & 1  &      & NF87,M06       \\
\qq & J004700+425855 & 00:47:00.77 & +42:58:55.20 & 21.76$\pm$0.19 & 20.25$\pm$0.03 & 21.26$\pm$0.05 & 22.26$\pm$0.19  & 22.57$\pm$0.51 &  1  & 1  & SP   & NF87,M06       \\
\qq & J004700+404923 & 00:47:00.98 & +40:49:23.16 & 23.30$\pm$0.60 & 21.50$\pm$0.07 & 22.34$\pm$0.16 & 24.62$\pm$0.67  & 21.65$\pm$0.43 &  1  & 1  & NS   & HK04,M06       \\
\qq & J004706+420715 & 00:47:06.36 & +42:07:15.24 & 21.52$\pm$0.21 & 20.31$\pm$0.03 & 21.56$\pm$0.11 & 24.31$\pm$0.74  & 21.74$\pm$0.44 &  1  & 1  &      & M06            \\
\qq & J004711+423047 & 00:47:11.93 & +42:30:47.16 & 23.01$\pm$0.58 & 20.94$\pm$0.04 & 21.65$\pm$0.09 & 23.34$\pm$0.55  & 22.64$\pm$0.72 &  1  & 1  &      & NF87,M06       \\
\qq & J004725+425859 & 00:47:25.90 & +42:58:59.88 & 23.37$\pm$0.78 & 20.97$\pm$0.04 & 21.87$\pm$0.11 & 23.01$\pm$0.41  & 22.96$\pm$0.83 &  1  & 1  & SP   & M06            \\
\qq & J004730+430341 & 00:47:30.34 & +43:03:41.04 & 23.18$\pm$0.57 & 20.41$\pm$0.03 & 21.55$\pm$0.09 & 22.92$\pm$0.44  & 22.31$\pm$0.51 &  1  & 1  & GS   & M06            \\
\qq & J004731+422618 & 00:47:31.58 & +42:26:18.24 & 23.17$\pm$0.76 & 21.50$\pm$0.06 & 22.35$\pm$0.16 & 22.71$\pm$0.32  & 23.03$\pm$0.67 &  1  & 0  &      & M06            \\
\qq & J004732+421135 & 00:47:32.93 & +42:11:35.88 & 22.00$\pm$0.28 & 20.43$\pm$0.03 & 21.08$\pm$0.06 & 21.57$\pm$0.16  & 23.03$\pm$0.64 &  1  & 1  &      & M06            \\
\qq & J004734+420408 & 00:47:34.15 & +42:04:08.04 & 22.50$\pm$0.44 & 20.88$\pm$0.04 & 21.49$\pm$0.09 & 22.95$\pm$0.48  & 23.07$\pm$0.65 &  1  & 4  &      &                \\
\qq & J004746+421214 & 00:47:46.75 & +42:12:14.76 & 21.92$\pm$0.25 & 20.40$\pm$0.03 & 21.71$\pm$0.11 & 22.50$\pm$0.35  & 22.56$\pm$0.68 &  1  & 1  &      &                \\
\qq & J004753+421457 & 00:47:53.59 & +42:14:57.48 & 22.26$\pm$0.38 & 21.06$\pm$0.04 & 22.05$\pm$0.15 & 22.89$\pm$0.44  & 23.76$\pm$0.83 &  1  & 0  &      & M06            \\
\qq & J004757+412810 & 00:47:57.38 & +41:28:10.56 & 23.83$\pm$1.01 & 22.17$\pm$0.10 & 22.58$\pm$0.24 & 22.84$\pm$0.36  & 22.32$\pm$0.56 &  2  & 0  &      &                \\
\qq & J004758+430006 & 00:47:58.61 & +43:00:06.48 & 21.94$\pm$0.26 & 20.56$\pm$0.03 & 21.21$\pm$0.19 & 22.88$\pm$0.48  & 22.80$\pm$0.86 &  1  & 1  & SP   & NF87,M06       \\
\qq & J004801+414410 & 00:48:01.20 & +41:44:10.68 & 23.21$\pm$0.84 & 22.03$\pm$0.19 & 22.54$\pm$0.50 & 24.20$\pm$0.68  & 21.06$\pm$0.41 &  2  & 3  &      &                \\
\qq & J004803+423637 & 00:48:03.53 & +42:36:37.80 & 21.97$\pm$0.24 & 20.70$\pm$0.03 & 21.40$\pm$0.06 & 22.37$\pm$0.20  & 22.39$\pm$0.65 &  1  & 1  &      & NF87,M06       \\
\qq & J004804+423510 & 00:48:04.90 & +42:35:10.68 & 23.16$\pm$0.65 & 21.79$\pm$0.06 & 22.88$\pm$0.22 & 24.33$\pm$0.53  & 23.62$\pm$0.50 &  2  & 1  &      & M06            \\
\qq & J004811+414453 & 00:48:11.02 & +41:44:53.88 & 22.13$\pm$0.35 & 21.46$\pm$0.05 & 21.86$\pm$0.11 & 22.27$\pm$0.26  & 21.52$\pm$0.42 &  2  & 0  &      &                \\
\qq & J004822+404541 & 00:48:22.15 & +40:45:41.40 & 22.18$\pm$0.21 & 21.02$\pm$0.04 & 21.68$\pm$0.08 & 23.29$\pm$0.42  & 22.99$\pm$0.52 &  1  & 1  & NS   & HK04,M06       \\
\qq & J004824+410811 & 00:48:24.79 & +41:08:11.04 & 21.76$\pm$0.20 & 19.80$\pm$0.03 & 21.00$\pm$0.05 & 22.25$\pm$0.23  & 22.26$\pm$0.68 &  2  & 1  & NS   & HK04           \\
\qq & J004830+414700 & 00:48:30.41 & +41:47:00.96 & 23.27$\pm$0.82 & 21.06$\pm$0.04 & 22.52$\pm$0.20 & 24.30$\pm$0.69  & 21.86$\pm$0.51 &  1  & 1  &      &                \\
\qq & J004847+425136 & 00:48:47.90 & +42:51:36.00 & 22.11$\pm$0.28 & 20.34$\pm$0.02 & 20.92$\pm$0.04 & 22.17$\pm$0.17  & 21.01$\pm$0.27 &  1  & 1  &      & M06            \\
\qq & J004854+424953 & 00:48:54.05 & +42:49:53.76 & 22.49$\pm$0.38 & 21.00$\pm$0.04 & 21.92$\pm$0.09 & 22.50$\pm$0.22  & 23.25$\pm$0.58 &  1  & 0  &      & NF87,M06       \\
\qq & J004901+415245 & 00:49:01.10 & +41:52:45.48 & 22.12$\pm$0.32 & 21.25$\pm$0.05 & 21.69$\pm$0.10 & 22.10$\pm$0.30  & 22.33$\pm$0.57 &  2  & 0  &      &                \\
\qq & J004915+412046 & 00:49:15.70 & +41:20:46.68 & 22.34$\pm$0.30 & 20.13$\pm$0.02 & 21.45$\pm$0.07 & 22.24$\pm$0.24  & 23.20$\pm$0.69 &  1  & 1  & NS   &                \\
\qq & J004931+423737 & 00:49:31.94 & +42:37:37.92 & 22.38$\pm$0.25 & 20.72$\pm$0.03 & 21.58$\pm$0.09 & 22.53$\pm$0.28  & 23.44$\pm$0.56 &  1  & 1  &      & M06            \\
\qq & J004949+423139 & 00:49:49.82 & +42:31:39.72 & 23.10$\pm$0.51 & 21.19$\pm$0.04 & 21.93$\pm$0.10 & 22.94$\pm$0.28  & 22.25$\pm$0.42 &  1  & 1  & GS   &                \\
\qq & J005038+411815 & 00:50:38.66 & +41:18:15.84 & 22.55$\pm$0.42 & 20.66$\pm$0.03 & 21.76$\pm$0.08 & 23.19$\pm$0.41  & 22.29$\pm$0.92 &  1  & 1  & NS   &                \\
\qq & J005123+435321 & 00:51:23.45 & +43:53:21.84 & 22.36$\pm$0.30 & 20.80$\pm$0.03 & 21.83$\pm$0.09 & 22.87$\pm$0.32  & 22.43$\pm$0.48 &  1  & 1  & NE   &                \\
\qq & J005130+420657 & 00:51:30.62 & +42:06:57.96 & 22.93$\pm$0.52 & 21.08$\pm$0.04 & 21.78$\pm$0.09 & 22.66$\pm$0.24  & 21.71$\pm$0.43 &  1  & 1  & NS   &                \\
\qq & J005134+415704 & 00:51:34.99 & +41:57:04.68 & 23.55$\pm$0.65 & 21.24$\pm$0.07 & 21.75$\pm$0.11 & 23.58$\pm$0.68  & 22.32$\pm$0.63 &  1  & 1  & NS   &                \\
\qq & J005200+430323 & 00:52:00.12 & +43:03:23.76 & 22.70$\pm$0.39 & 21.41$\pm$0.05 & 22.08$\pm$0.12 & 24.04$\pm$0.65  & 22.46$\pm$0.60 &  1  & 0  &      & NF87,M06       \\
\qq & J005232+440613 & 00:52:32.04 & +44:06:13.32 & 23.21$\pm$0.70 & 21.26$\pm$0.04 & 21.75$\pm$0.09 & 23.00$\pm$0.38  & 22.23$\pm$1.55 &  1  & 0  & NE   &                \\
\qq & J005247+442257 & 00:52:47.50 & +44:22:57.72 & 21.95$\pm$0.24 & 21.02$\pm$0.04 & 21.82$\pm$0.09 & 22.79$\pm$0.31  & 22.02$\pm$1.49 &  2  & 1  & NE   &                \\
\qq & J005357+422928 & 00:53:57.19 & +42:29:28.32 & 25.16$\pm$0.63 & 20.11$\pm$0.02 & 24.22$\pm$0.58 & 24.49$\pm$0.67  & 19.63$\pm$0.07 &  1  & 0  &      &                \\
\qq & J010956+482118 & 01:09:56.42 & +48:21:18.72 & 22.67$\pm$1.26 & 20.93$\pm$0.07 & 21.41$\pm$0.13 & 21.91$\pm$0.29  & 20.90$\pm$0.33 &  1  & 0  &      &
\enddata
\tablenotetext{a}{1 -- first priority candidate; 2 -- second priority candidate}
\tablenotetext{b}{0 -- not observed; 1 -- confirmed PN; 2 -- not detected in our follow-up,
but identified later via cross-identifications with other catalogs;
3 -- not detected in our follow-up;
4 -- our unobserved candidates with obvious emission in [O\,{\sc iii}] $\lambda$5007 and H$\alpha$ narrow filters from SLGG data}
\tablenotetext{c}{SW -- SW Shell;
		  G1 -- G1 Clump;
		  GL -- G1 Clump loop;
		  LO -- NGC\,205 Loop;
		  NGC205 -- in NGC\,205;
		  SP -- Northern Spur;
		  GS -- Giant Stream;
		  NS -- NE Shelf;
		  NE -- Andromeda NE
}
\tablenotetext{d}{Cross-identification from other catalogues: 
JF86 -- \citet{JF86}; NF87 -- \citet{NF87}; HK04 -- \citet{HK04}; 
M06 -- \citet{Mer06}.}
\end{deluxetable}

\setcounter{qub}{0}
%
%...Heliocentric velocities for detected PNe in M31
%
\begin{deluxetable}{rllllllc}
%\rotate
%\tabletypesize{\tiny}
%\tabletypesize{}
\tablecaption{Heliocentric velocities and observed line fluxes of our
	      confirmed PNe in M31 \label{tbl-2}}
\tablewidth{0pt}
\tablehead{
\colhead{\#} &\colhead{SDSS name} & \colhead{V$\pm$dV} & \colhead{F(H$\beta$)$^a$}&
\colhead{F(5007)$^a$} & \colhead{F(H$\alpha$)$^a$} & \colhead{F(6584)$^a$} & \colhead{C(H$\beta$)} \\
 & & \colhead{km/s} &                &                &                &                & \\
&\colhead{(1)} & \colhead{(2)} & \colhead{(3)} & \colhead{(4)} &  \colhead{(5)} &
\colhead{(6)} & \colhead{(7)}
}
\startdata
\qq&  J003447+393612 & $-$440$\pm$16 &    6.2 &   89.5 &   27.5  &    3.2 &    0.56    \\
\qq&  J003451+395429 & $-$413$\pm$11 &    5.4 &   77.4 &   17.9  &    6.5 &    0.17    \\
\qq&  J003451+402254 & $-$478$\pm$10 &\nodata &  132.0 &   27.1  &    5.6 & \nodata    \\
\qq&  J003533+405521 & $-$374$\pm$10 &   15.8 &   99.4 &   47.5  &    5.2 &    0.08    \\
\qq&  J003559+414050 & $-$103$\pm$17 &\nodata &   27.9 &   10.4  &\nodata & \nodata    \\
\qq&  J003612+393541 & $-$411$\pm$13 &   10.7 &  120.2 &   36.2  &\nodata &    0.21    \\
\qq&  J003628+393526 & $-$445$\pm$8  &   21.5 &  368.8 &   68.6  &    8.1 &    0.11    \\
\qq&  J003635+401737 & $-$354$\pm$31 &\nodata &   28.6 &    7.1  &\nodata & \nodata    \\
\qq&  J003707+383856 & $-$327$\pm$10 &\nodata &  107.1 &   24.1  &\nodata & \nodata    \\
\qq&  J003717+402437 & $-$457$\pm$31 &    4.0 &   60.2 &   16.6  &    4.5 &    0.47    \\
\qq&  J003721+404142 & $-$451$\pm$21 &\nodata &   22.7 &    5.8  &\nodata & \nodata    \\
\qq&  J003736+402802 & $-$395$\pm$10 &   10.5 &  116.4 &   36.5  &    5.0 &    0.24    \\
\qq&  J003739+393009 & $-$423$\pm$10 &    4.5 &   49.3 &   16.4  &\nodata &    0.30    \\
\qq&  J003749+405354 & $-$305$\pm$13 &   15.6 &  146.6 &   65.3  &    8.2 &    0.50    \\
\qq&  J003751+404547 & $-$525$\pm$12 &   18.4 &  185.2 &   71.5  &\nodata &    0.40    \\
\qq&  J003841+394739 & $-$511$\pm$23 &    3.0 &   31.2 &   11.9  &\nodata &    0.42    \\
\qq&  J003846+412819 & $-$228$\pm$6  &    7.7 &   93.8 &   32.1  &\nodata &    0.48    \\
\qq&  J003848+413937 & $-$223$\pm$11 &    4.8 &   52.9 &   16.4  &    2.8 &    0.22    \\
\qq&  J003855+410655 & $-$421$\pm$11 &\nodata &   82.2 &   16.6  &\nodata & \nodata    \\
\qq&  J003901+415111 & $-$202$\pm$8  &   16.3 &  143.7 &   51.2  &\nodata &    0.12    \\
\qq&  J003902+402250 & $-$507$\pm$12 &    6.4 &   76.6 &   19.8  &    3.8 &    0.09    \\
\qq&  J003922+410657 & $-$381$\pm$8  &   12.1 &  187.1 &   41.7  &   11.1 &    0.22    \\
\qq&  J003945+403142 & $-$553$\pm$12 &   10.9 &  200.0 &   38.5  &   16.5 &    0.24    \\
\qq&  J003948+411043 & $-$435$\pm$14 &    3.1 &   49.2 &   12.7  &\nodata &    0.45    \\
\qq&  J003956+410507 & $-$297$\pm$15 &    9.2 &  100.6 &   33.4  &   10.5 &    0.29    \\
\qq&  J004000+410409 & $-$312$\pm$12 &\nodata &  149.5 &   29.1  &\nodata & \nodata    \\
\qq&  J004035+421031 & $-$114$\pm$16 &    6.1 &   42.0 &   19.0  &    4.3 &    0.14    \\
\qq&  J004056+402008 & $-$556$\pm$10 &    8.7 &  143.0 &   27.8  &    6.5 &    0.11    \\
\qq&  J004212+423134 & $-$305$\pm$12 &    4.4 &   58.3 &   15.0  &    3.3 &    0.22    \\
\qq&  J004252+402903 & $-$278$\pm$11 &    4.0 &   71.4 &   14.8  &    5.2 &    0.29    \\
\qq&  J004258+424732 & $-$106$\pm$15 &    6.3 &   46.5 &   26.7  &\nodata &    0.51    \\
\qq&  J004259+420213 & $-$316$\pm$15 &    4.4 &   42.2 &   16.1  &\nodata &    0.34    \\
\qq&  J004311+422045 & $-$268$\pm$11 &\nodata &   92.7 &   22.3  &\nodata & \nodata    \\
\qq&  J004327+424203 & $-$136$\pm$12 &   10.1 &  118.6 &   35.3  &\nodata &    0.25    \\
\qq&  J004332+415842 & $-$587$\pm$15 &    6.4 &   92.8 &   32.4  &    8.3 &    0.72    \\
\qq&  J004342+422235 & $-$201$\pm$14 &    4.0 &   59.7 &   14.8  &    2.7 &    0.31    \\
\qq&  J004402+421716 & $-$259$\pm$12 &    8.4 &  115.6 &   34.6  &    6.6 &    0.46    \\
\qq&  J004403+422746 & $-$211$\pm$10 &   12.0 &  144.4 &   41.5  &    7.6 &    0.24    \\
\qq&  J004454+421738 & $-$227$\pm$17 &    5.4 &   54.9 &   17.8  &    5.4 &    0.17    \\
\qq&  J004542+425526 & $-$126$\pm$12 &   11.9 &  179.5 &   39.5  &    8.4 &    0.17    \\
\qq&  J004552+423652 & $-$162$\pm$9  &   18.4 &  181.3 &   57.8  &   11.4 &    0.12    \\
\qq&  J004613+424028 &  $-$96$\pm$12 &    8.0 &  101.9 &   23.7  &\nodata &    0.03    \\
\qq&  J004626+430043 & $-$418$\pm$16 &   14.6 &  225.4 &   47.2  &   16.7 &    0.14    \\
\qq&  J004642+420835 &    ~~6$\pm$10 &   15.7 &  308.9 &   48.2  &    9.6 &    0.05    \\
\qq&  J004652+431058 & $-$425$\pm$14 &    7.5 &  133.8 &   26.3  &    5.1 &    0.22    \\
\qq&  J004655+422401 & $-$124$\pm$15 &   12.1 &  238.3 &   53.9  &   14.7 &    0.54    \\
\qq&  J004659+423758 &  $-$79$\pm$12 &   11.5 &  100.5 &   33.8  &   12.5 &    0.05    \\
\qq&  J004700+425855 &  $-$21$\pm$33 &    7.2 &  137.6 &   31.0  &\nodata &    0.49    \\
\qq&  J004700+404923 & $-$348$\pm$11 &    5.0 &   59.1 &   17.4  &    6.9 &    0.25    \\
\qq&  J004706+420715 &  $-$43$\pm$6  &   14.1 &  227.4 &   42.3  &    9.1 &    0.04    \\
\qq&  J004711+423047 &  $-$88$\pm$8  &    9.1 &   93.1 &   29.2  &   12.5 &    0.16    \\
\qq&  J004725+425859 & $-$163$\pm$26 &\nodata &   63.9 &   22.3  &\nodata & \nodata    \\
\qq&  J004730+430341 & $-$438$\pm$11 &    8.5 &   93.2 &   29.0  &\nodata &    0.22    \\
\qq&  J004732+421135 & $-$146$\pm$8  &    8.9 &  105.4 &   35.3  &    8.5 &    0.41    \\
\qq&  J004746+421214 &  $-$86$\pm$12 &\nodata &  105.4 &   33.4  &\nodata & \nodata    \\
\qq&  J004758+430006 &  $-$85$\pm$15 &    8.5 &   48.0 &   28.1  &\nodata &    0.22    \\
\qq&  J004803+423637 &  $-$78$\pm$15 &    6.5 &   70.6 &   29.3  &\nodata &    0.58    \\
\qq&  J004804+423510 &  $-$80$\pm$30 &\nodata &   12.6 &    6.8  &\nodata & \nodata    \\
\qq&  J004822+404541 & $-$309$\pm$11 &    6.1 &   50.2 &   19.2  &\nodata &    0.14    \\
\qq&  J004824+410811 & $-$364$\pm$15 &   21.5 &  362.4 &   84.6  &   13.9 &    0.38    \\
\qq&  J004830+414700 & $-$127$\pm$15 &    3.3 &   75.7 &   12.9  &\nodata &    0.34    \\
\qq&  J004847+425136 &    ~~8$\pm$27 &\nodata &  122.2 &   39.3  &\nodata & \nodata    \\
\qq&  J004915+412046 & $-$350$\pm$20 &   13.5 &  207.7 &   48.8  &    5.9 &    0.28    \\
\qq&  J004931+423737 & $-$106$\pm$20 &   10.5 &  133.8 &   33.5  &\nodata &    0.12    \\
\qq&  J004949+423139 & $-$367$\pm$14 &    4.5 &   69.6 &   19.1  &\nodata &    0.48    \\
\qq&  J005038+411815 & $-$170$\pm$20 &    6.3 &  104.9 &   20.7  &\nodata &    0.14    \\
\qq&  J005123+435321 & $-$135$\pm$12 &    7.3 &  108.5 &   24.2  &    7.1 &    0.17    \\
\qq&  J005130+420657 & $-$222$\pm$10 &    9.0 &  111.8 &   35.0  &   11.4 &    0.38    \\
\qq&  J005134+415704 & $-$339$\pm$13 &    7.2 &   75.5 &   22.1  &    7.2 &    0.09    \\
\qq&  J005247+442257 & $-$129$\pm$12 &    6.1 &   66.6 &   20.5  &    4.5 &    0.20
\enddata
\tablenotetext{a}{Observed flux in units of 10$^{-16}$ ergs\ s$^{-1}$cm$^{-2}$}
\end{deluxetable}

\end{document}